\documentclass[12pt,letterpaper]{JHEP3}
\pdfoutput=1
\usepackage{epsfig}
\usepackage{subcaption}
\usepackage{graphicx}
\usepackage[english]{babel}

\usepackage{cite}

\def\be{\begin{equation}}
\def\ee{\end{equation}}
\def\rads{R_{\rm AdS}}
\def\rs{R_{\rm BH}}
\def\lp{l_{\rm Planck}}
\def\be{\begin{equation}}
\def\ee{\end{equation}}
\def\sbh{S_{\rm BH}}

\graphicspath{{Figures/}}

\title{Position space analysis of the AdS (in)stability problem}

\author{Fotios V. Dimitrakopoulos, Ben Freivogel, Matthew Lippert, and I-Sheng Yang\\
ITFA and GRAPPA, Universiteit van Amsterdam, \\
Science Park 904, 1090 GL Amsterdam, Netherlands
}

\abstract{We investigate whether arbitrarily small perturbations in global AdS space are generically unstable and collapse into black holes on the time scale set by gravitational interactions. We argue that current evidence, combined with our analysis, strongly suggests that a set of nonzero measure in the space of initial conditions does not collapse on this time scale. We perform an analysis in position space to study this puzzle, and our formalism allows us to directly study the vanishing-amplitude limit.  We show that gravitational self-interaction leads to tidal deformations which are equally likely to focus or defocus energy, and we sketch the phase diagram accordingly. We also clarify the connection between gravitational evolution in global AdS and holographic thermalization.
}

\begin{document}

\section{Introduction}

Recently, the gravitational (in)stability of classical gravity in asymptotically global AdS spacetime has drawn a lot of attention and also generated a lot of confusion. Given a spherically symmetric perturbation of arbitrarily small initial amplitude $\epsilon$, two dramatically different behaviors have been observed at the timescale $\sim \epsilon^{-2}$, the earliest time on which interactions can have a significant effect\cite{BizRos11,deOPan12,Lie12,DiaHor12,Mal12,BucLeh12,BucLie13,Biz13,MalRos13,MalRos13a,BalBuc14,MalRos14,HorSan14}. Sometimes a black hole forms around this time; sometimes a long-lived quasi-periodic behavior emerges and gravity does not become strong. This is a great puzzle concerning both the gravitational dynamics in the bulk and the corresponding thermalization process in the holographic boundary theory.

In this paper we will focus on the bulk perspective and on the simple case of a free massless scalar field coupled to gravity. We treat the system classically and impose spherical symmetry.  In the limit of small amplitude $\epsilon$, the energy density is proportional to $\epsilon^2$ and controls the strength of gravitational effects. Therefore, behavior at the time scale $\epsilon^{-2}$ is sensitive to the leading-order effects of gravitational interactions. One framework to study this is to analyze the nonlinear couplings between the linearized modes induced by the gravitational interactions. Linearized modes in AdS space all have frequencies which are integer multiples of the AdS scale.  A mode that is initially unexcited can be resonantly driven by the excited modes, which allows for the possibility of efficient transfer of energy. Such efficient energy transfer between modes generically leads to the breakdown of na\"ive perturbation theory, since the true solution does not remain close to the solution in the non-interacting theory. This resonance effect was argued to be the cause of an energy cascade---energy spreads out into more and higher modes---in order to explain black hole formation and the power-law spectrum observed during such processes\cite{BizRos11,DiaHor11,DiaHor12,deOPan12}. It was also argued that since the AdS spectrum is resonant, such an instability should be the generic outcome of small perturbations.

Counter-examples to the above claim in the form of the stable, quasi-periodic solutions were initially viewed as being special. It was conjectured in \cite{DiaHor12} that these stable solutions will shrink to a set of measure zero in the small $\epsilon$ limit, and the term ``stability island'' was used to describe their existence in the generically unstable sea of phase space. However, more recent evidence suggests that such a conclusion is too strong.\footnote{I.S. Yang thanks Jorge Santos for stressing this point during a discussion.} Numerical evidence suggests that, at finite $\epsilon$, the stable and unstable solutions both have nonzero measure in the space of initial conditions \cite{BucLie13,MalRos13,BalBuc14}.\footnote{Note that we are always discussing stability on the interaction time scale $\epsilon^{-2}$ in this paper. The question of the behavior on longer time scales is a fascinating one that touches on issues of ergodicity, Arnold diffusion, and the KAM theorem. We do not know how to attack questions on these longer time scales analytically or numerically.} One can then apply a simple scaling argument, described in more detail in Sec.~\ref{sec-review}, to show that in the $\epsilon\rightarrow0$ limit, the stable solutions persist.  However, the same argument fails for unstable solutions. The open question now becomes whether there are ``instability corners.'' Namely, in the $\epsilon\rightarrow0$ limit, do the unstable solutions shrink to a set of measure zero, or do also continue to have finite measure?

There are some important misconceptions and misunderstandings in the current literature regarding the status of the AdS (in)stability problem, due in part to three points of confusion, which we would like to clarify here. First of all, an energy cascade is not identical to, nor does it guarantee black hole formation. This distinction has not been made clear enough. Both have been frequently used interchangeably and referred to as the ``instability of AdS space.''  Black hole formation requires energy to be focused into a small spatial region. According to the uncertainty principle, energy flowing to high momentum is certainly a necessary condition for that, but it is not sufficient. It is entirely possible for even unboundedly high momentum modes to be populated, but for the energy distribution to stay roughly spatially homogeneous. 

Therefore, in this paper, {\bf AdS instability} strictly refers to {\bf black hole formation} only. Because the AdS geometry changes dramatically in this case, such nomenclature aligns with a more gravity point of view.\footnote{From the hydrodynamic point of view, the existence of an energy cascade might be a suitable definition of instability. Indeed, this is the perspective taken by some authors, and we wish the reader to see the distinction clearly.} This also allows us to study its implications on the boundary CFT. When we refer to a solution or initial condition as stable or unstable, we will always be indicating whether it collapses to form black hole or not.

The second point of confusion is the use of term ``generic.''
Numerical evidence suggests that, at finite $\epsilon$, the stable and unstable solutions both form sets of nonzero measure in the space of initial conditions\cite{BucLie13,MalRos13,BalBuc14}.\footnote{Strictly speaking, numerical results only cover discrete choices of initial conditions.  So, it is therefore impossible on numerical grounds alone to prove that any such set has either zero or nonzero measure. 
This fact holds equally for both stable and unstable solutions. Nevertheless, if either set really were measure-zero, unless the numerical code secretly enforced extra symmetries, it would be extremely unlikely to find such a result even once in simulations. Thus, despite the numerical controversy over some of the stable solutions \cite{BizRos14}, we still interpret the current evidence that stable and unstable solutions both have nonzero measures.} We are interested in the $\epsilon\rightarrow0$ limit, and in this paper we use the following definition:
\begin{enumerate}
\item ``Generic instability'' means the set of stable initial conditions (not forming black holes) shrinks to measure zero.
\item ``Generic stability'' means the set of unstable initial conditions (forming black holes) shrinks to measure zero.
\item ``Mixed'' means that both sets have nonzero measure as $\epsilon\rightarrow0$.
\end{enumerate}
Until recently, references in the literature did not clearly distinguish between (1) and (3). For example, it was conjectured in\cite{DiaHor12} that ``stability islands'' shrink to a set of measure zero in the small-$\epsilon$ limit, which is certainly arguing for only (1). However, the numerical evidence in \cite{BizRos11} showing that black holes continue to form as $\epsilon$ is reduced is consistent with both (1) and (3). Since these are three physically different cases, we think such a clear distinction is needed.

Finally, when addressing the question of instability, one needs to specify a time scale. In this paper, we will only discuss the time scale that goes to infinity as $\epsilon^{-2}$ in the $\epsilon\rightarrow0$ limit.\footnote{Behaviors at shorter time scales are somewhat trivial. For example, for a given fixed time, black hole forming solutions disappear as $\epsilon\rightarrow0$, so the system is generically stable, case (2). The behavior at longer time scales is a very deep problem that touches on issues of ergodicity, Arnold diffusion, and the KAM theorem.  We do not know how to attack those questions analytically or numerically.} Indeed a na\"ive perturbation analysis shows that something interesting can happen at this time scale. The physical question we will address is whether that ``something interesting'' is generically black hole formation? In the end, we will try to relate the answer to the boundary CFT: Does the boundary system thermalize at this time scale?

After making all these definitions clear, in Sec.~\ref{sec-review} we briefly review the recent progress on this topic. We then present a very simple scaling argument which shows that possibility (1) defined above, ``generic instability'', is the most unlikely given by existing evidence.\footnote{I.S. Yang thanks Jorge Santos for stressing this point during a discussion.} This directly argues against the ``stability island'' conjecture\cite{DiaHor12}. The remaining question then is whether AdS space is generically stable (2) or mixed (3).

In Sec.~\ref{sec-weakG} we set up our perturbative method for studying gravitational self-interaction.  This position-space approach is more directly relevant than the usual momentum-space analysis to the question of whether or not black holes form.\footnote{In principle, one can include all the information about relative phases in the spectral analysis to achieve the same result. Our position-space approach is simply more direct. In addition, it technically circumvents the subtlety that the gravitational interaction imposes significant additional constraints on possible resonances \cite{CraEvn14}.} If energy gets focused into a smaller region, then the solution is evolving toward a black hole. If energy is defocused into a larger region, then the solution is evolving away from a black hole. We explicitly demonstrate that in the $\epsilon\rightarrow0$ limit, the focusing/defocusing dynamics depend only on the gravitational self-interaction near the origin of AdS, when the energy of the perturbation is maximally concentrated. The propagation through the rest of asymptotic AdS space plays no dynamical role.

\begin{figure}[tb]
\begin{center}
\includegraphics[width=12cm]{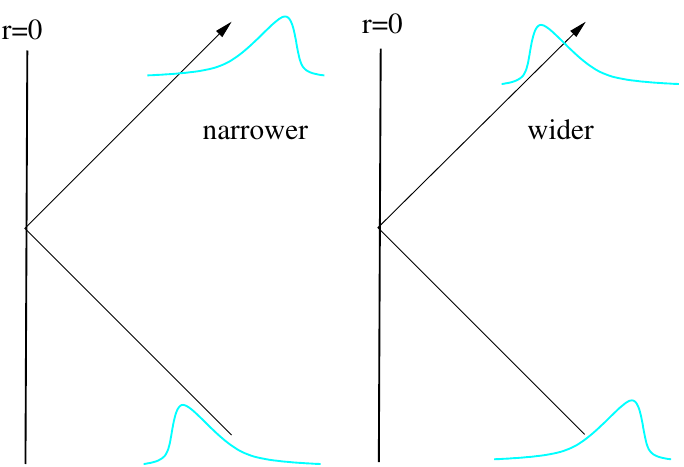}
\caption{A thin shell that has higher energy density in its front will come out narrower after gravitational self-interactions as it bounces through $r=0$. A shell with higher energy density in its tail will come out wider after the bounce.
\label{fig-flip}}
\end{center}
\end{figure}

In Sec.~\ref{sec-proof} we prove a one-to-one correspondence between focusing and defocusing energy in the near-center dynamics. Heuristically, our result is shown in Fig.~\ref{fig-flip}: A shell of massless scalar field will become narrower, its energy focused, if it is denser in the front. On the other hand, if it is denser in the tail, it will become wider and energy will defocus.\footnote{The shell profile will change in other ways, but all changes are suppressed by $\epsilon^2$. The focusing or defocusing behavior will last for a time scale $\lesssim\epsilon^{-2}$, so it is the dynamics we are interested in here.} More generally, the leading-order dynamics of focusing and defocusing are related by time reversal, so a local maximum of energy density is also equally likely to grow or diminish within the time scale $\lesssim\epsilon^{-2}$.

In Sec.~\ref{sec-dis} we present the new intuition our method provides and propose a conjecture on the structure of the phase space. Based on the symmetry between focusing and defocusing dynamics, the stable, quasi-periodic solutions can be understood as trajectories that alternate between the two. As a result, they may form quasi-closed loops in phase space. In fact, some unstable solutions are also known to exhibit this alternating behavior while in the weak-gravity regime. Based on this understanding of the dynamics, we propose a conjecture on how to visualize the phase space of small perturbations in AdS space. 

We also discuss how these gravitational calculations can shed light on the general concepts of thermalization in a closed system. In particular, contrary to conventional wisdom, black hole formation at the $\epsilon^{-2}$ time scale is not necessarily the holographic dual of thermalization in the boundary field theory.  If the thermal gas phase is the equilibrium state, then black hole formation describes prethermalization that significantly delays true thermalization\cite{TroChe11,GriKuh11}.

In Sec.~\ref{sec-sum}, we provide a quick summary of six major points of this paper. In Appendices \ref{sec-shift} and \ref{sec-examples}, we provide the computational details of our method and numerical examples to demonstrate how the shape of the profile determines whether its energy is focused or defocused.


\section{Stability islands or instability corners?}
\label{sec-review}


We are interested in the perturbative stability of global AdS space. We will work in $(3+1)$ dimensions and employ the following metric for vacuum $AdS_4$: 
\be
\label{AdSmetric}
ds_{AdS_4}^2 =  -\left(1+ \frac{r^2}{\rads^2} \right) dt^2 + \frac{dr^2}{1+\frac{r^2}{\rads^2}} + r^2 d\Omega_2^2 
\ee
where $\rads$ is the AdS radius and $d\Omega_2^2$ is the round metric on $S^2$.\footnote{Our radial coordinate $r$ is related to the radial coordinate $x$ used in \cite{BizRos11} by $r = \tan x$.}  Our perturbations will take the form of a real, massless scalar field $\phi$ minimally coupled to Einstein gravity with a negative cosmological constant:
\be
\label{fullaction}
S = \int d^4x \  \sqrt{g} \left( \frac{1}{16\pi} R + \frac{6}{\rads^2} - \frac{1}{2} \partial_\mu \phi \partial^\mu \phi \right) \ .
\ee
The Planck scale has been set to one. We will consider only spherically symmetric solutions, so both the scalar $\phi$ and the metric functions $g_{tt}$ and $g_{rr}$ will only depend on $t$ and $r$.

Here we will review some existing evidence and argue that a careful interpretation strongly supports the following conclusion for spherically symmetric perturbations of a massless scalar field in AdS space: \\ \ \\
{\it In the $\epsilon\rightarrow0$ limit, at the $T\sim\epsilon^{-2}$ time scale,  AdS space is either generically stable, or that neither stable nor unstable perturbations are generic.} \\

The first part of our argument is based on ample numerical evidence at small but finite $\epsilon$.  The initial conditions that lead to black hole formation (unstable) and those that lead to quasi-periodic solutions (stable) both form open sets in the phase space of nonzero measure. Note that the phase space is infinite dimensional, so no numerical evidence can prove that any set is really open. Nevertheless, whatever extrapolations are being made should be applied equally to both stable and unstable solutions, and the existing numerical evidence is quite sufficient to show that they are on equal footing. More specifically, numerical tests can scan a one-parameter family of initial conditions, corresponding to a line in phase space. It has been clearly demonstrated that for a few  such lines, the initial conditions that lead to stable and unstable solutions both form finite segments \cite{BucLie13,MalRos13,BalBuc14}. We will pragmatically take this as evidence that both stable and unstable sets in phase space have nonzero measure at small but finite $\epsilon$.

In particular, within the set of stable solutions, one can identify a subset for which ``gravity never becomes strong'' during the $\sim\epsilon^{-2}$ time scale; that is, 
\begin{equation}
\exists~\phi(\epsilon,r,t)~,~ {\rm such  \ that \ } 
\left(\dot\phi^2+\phi'^2\right)<\delta\ll1~ {\rm for \ } 0\leq t\leq T\sim\epsilon^{-2}.
\end{equation}
Our next step is to show that in the $\epsilon\rightarrow0$ limit, these stable solutions cannot disappear.  We can use the scaling behavior observed in \cite{BizRos11,BalBuc14}, which was trustworthy to leading order in $\epsilon$. We will demonstrate that in the $\epsilon\rightarrow0$ limit, this scaling behavior is exact for stable, weak-gravity solutions. 

The spectrum of a massless field in the AdS background is given by integer multiples of the AdS energy scale $\rads^{-1}$, meaning that the field profile is exactly periodic in time.  Heuristically, a spherical wavefront shrinks toward the origin $r=0$, passes through it, expands again to infinity, and finally bounces off the boundary back to the original position.\footnote{The periodicity of geodesics in AdS is $2\pi\rads$, and in that time they pass through the origin twice.  However, a shell of massless scalar field with Dirichlet boundary conditions at the boundary is actually periodic in half that time, $\pi\rads$, during which the wavefront passes through the origin only once.}   It is natural to describe the dynamics as a function of the ``number of bounces'' $N = \frac{t}{\pi \rads}$ instead of the microscopic time $t$:
\begin{equation}
\phi(r,N+1)\equiv\phi(r,t+\pi \rads) 
= \phi(r,N)\equiv\phi(r,t)~.
\end{equation}
Now, introducing gravitational self-interaction, as long as the field amplitude (and therefore the resulting back-reaction) is small, we have a small correction to the above exactly periodic solution,
\begin{equation}
\phi(r,N+1) - \phi(r,N) = 
A[\phi,\dot{\phi}]+\mathcal{O}(\phi^5)~.
\label{eq-onebounce}
\end{equation}
The functional $A$ describes the small, leading-order changes to the profile, which we will analyze further in the following sections. Here we only need to know that it scales like $\phi^3$. It is convenient to introduce the rescaled field, $\bar\phi\equiv \phi/\epsilon$, whose evolution is given by
\begin{equation}
\bar\phi(r,N+1) - \bar\phi(r,N) = 
A[\bar{\phi},\dot{\bar\phi}]\epsilon^2+\mathcal{O}(\epsilon^4)~,
\label{eq-diffscale}
\end{equation}
Although the value of $N$ is discrete, in the $\epsilon\rightarrow 0$ limit, the change due to each bounce goes to zero.  We can therefore take the continuum limit, in which Eq.~(\ref{eq-diffscale}) becomes 
\begin{equation}
\frac{d\bar\phi}{d(\epsilon^2N)} =  A~. 
\label{eq-contscale}
\end{equation} 
Thus, the scaling behavior is exact:
\begin{equation}
\bar\phi_\epsilon(r,N) = \bar\phi_{\frac{\epsilon}{\alpha}}(r,\alpha^2 N)~.
\end{equation}
Reducing the amplitude of the fluctuation simply slows down the dynamics by $\alpha^{2}$: if $\epsilon$ is reduced by a factor of $\alpha$, it takes $\alpha^2$ more bounces to reach the same configuration.  Therefore, if there is a stable solution at some finite $\epsilon$ and for a time $T\sim\epsilon^{-2}$ during which gravity never becomes strong, this must also be a stable solution at any smaller $\epsilon$, all the way to the $\epsilon\rightarrow0$ limit.\footnote{We should note that the expansion in powers of $\epsilon$ is most likely asymptotic\cite{Ren90}, but its leading-order result has been accurate for many similar applications\cite{BizChm08}.}

Interestingly, this same argument is not applicable to unstable solutions. In order to form a black hole, the scalar field profile must first evolve to have large energy density somewhere,
$\left(\dot\phi^2+\phi'^2\right)\sim1$. In other words, gravity must become strong, at which point the higher order terms in Eq.~(\ref{eq-diffscale}) cannot be ignored. In those cases the scaling behavior is lost. A collapsing solution at some small but finite $\epsilon$ might escape that fate if we reduce $\epsilon$ further\cite{BasKri14}.

At this point, we are left with two possibilities:
\begin{itemize}
\item Neither stable nor unstable perturbations are generic, since they both occupy sets of nonzero measure in the phase space.
\item AdS space is perturbatively stable generically, but there are special ``instability corners'', which shrink to measure zero in the limit $\epsilon \rightarrow 0$.
\end{itemize}
Finally, recall that we have so far limited ourselves to spherical symmetry. Intuitively, spherical symmetry arranges for matter to converge at the origin, which is helpful for gravitational collapse. So, even if the first of the above possibilities holds within spherical symmetry, it may be that without spherical symmetry the second is instead the case.

\section{Weak gravitational self-interaction in position space}
\label{sec-weakG}

\subsection{The two-region approximation}

We now present our approach to explicitly calculating the functional $A$ in Eq.~(\ref{eq-onebounce}). Our result, a precise expression for $A$, is given in Eq.~(\ref{Adef}). Many of its properties will help us to better understand the dynamics and the possibility of instability corners. Our calculation will be in position space. The advantage for this approach is easily seen if we first picture the evolution of a thin shell of total energy $E \sim\epsilon^2$, thickness $w$ and initial size $r_0$, such that $r_0\gg w$. This corresponds to an initial field profile that is roughly given by
\begin{equation}
\phi_0(r,t)|_{t\sim t_i} \sim \frac{-\epsilon\sqrt{w}}{r}~
f\left(-\frac{r-r_0+t-t_i}{w}\right)~.
\label{eq-init}
\end{equation}
We will take the profile $f(x)$ to be some function that peaks at $x=0$ and has compact support an order-one range around around this peak ({\it i.e.}~$f(x) = 0$ for $|x| \gtrsim 1$).\footnote{We choose the profile to have compact support only to make the subsequent calculations somewhat cleaner.  The shell only needs a narrow, well-defined width.  Alternately, one could take $f$ to have, for example, Gaussian tails without affecting the results.} Note that we have carefully chosen the dependence on $w$ such that it does not affect the total mass, which is controlled solely by $\epsilon$. The small-perturbation limit then corresponds to $\epsilon\rightarrow0$. 

Other papers studying similar scenarios choose various different initial conditions for the scalar field perturbation. Some authors take initial conditions that place the energy near $r=0$.  In other cases, the perturbation originates from a quench in the boundary CFT and appears as a wavefront coming in from $r=\infty$\cite{AbadaS14}. Remember that in the small-$\epsilon$ limit, the leading-order behavior is the same as in empty AdS space; that is, the radiation shell simply bounces back and forth between $r=0$ and $r=\infty$. Therefore, all of these initial positions of the shell are related by a shift in time on the order of $R_{\rm AdS}$. Since we are interested in the outcome at longer time scales, they are all equivalent for our purposes. 

One advantage of our position-space approach is that we can choose an $r_0$ which implements the following ``two-region'' approximation: 
\begin{itemize}
\item For $r<r_0$, we will ignore that the background is AdS space and consider only the back-reaction of the scalar field on Minkowski space. 
\item For $r>r_0$, we will ignore the scalar field back-reaction and treat the geometry as empty AdS space.
\end{itemize}
In order justify this simplification, we first recall the general form of the Schwarzschild-AdS metric:
\begin{equation}
ds_{SAdS_4}^2 = - \left(1 - \frac{2M(r)}{r} + \frac{r^2}{\rads^2} \right) dt^2 + \frac{dr^2}{1 - \frac{2M(r)}{r} + \frac{r^2}{\rads^2}} + r^2 d\Omega_2^2 
\end{equation}
where $M(r)$ is the total mass located inside the sphere of radius $r$.  

For $r<r_0$ we will ignore the $r^2/\rads^2 $ terms in $g_{tt}$ and $g_{rr}$ responsible for the AdS asymptotics and calculate $M(r)$ due to the back-reaction of the radiation shell. This effect is strongest when the shell is near the origin and its energy is concentrated in a small region within $r<w$. We find that $M(r) \sim \epsilon^2$.

At $r=r_0$, we will start including the AdS terms and ignoring the back-reaction terms, such that for $r>r_0$ the metric is just that of empty AdS space. Na\"ively, this is allowed if the metric at $r_0$ is approximately that of Minkowski space; that is, the corrections due to both AdS and back-reaction must be small:
\begin{equation}
\frac{r_0^2}{R_{\rm AdS}^2} \ll 1~ ~ ~ {\rm and} ~ ~ ~
\frac{\epsilon^2}{r_0}\ll1~.
\end{equation}
However, we should really ask for a stronger condition. 
 
Our perturbative back-reaction calculation will be organized as an expansion in powers of $\epsilon^2/w$, and we will work up to some power $n$ using the Minkowski background.  In order to be able to trust our results up to that order, we cannot allow the transition at $r_0$ to have a competing effect, meaning
\begin{equation}
\frac{r_0^2}{R_{\rm AdS}^2} \ll 
\left(\frac{\epsilon^2}{w}\right)^n~ ~ ~{\rm and} ~ ~ ~
\frac{\epsilon^2}{r_0}\ll
\left(\frac{\epsilon^2}{w}\right)^n~.
\label{eq-switchcond}
\end{equation}
For any $\rads$, we can choose the shell small enough and thin enough to accommodate the hierarchy of scales
\begin{equation}
R_{\rm AdS} \gg r_0 \gg w \gg \epsilon^2~,
\label{eq-hierarchy}
\end{equation}
which satisfies Eq.~(\ref{eq-switchcond}) for any choice of $n$. 

The two-region approximation provides a very simple picture.  In the $\epsilon\rightarrow0$ limit, the dynamical evolution is totally controlled by the central Minkowski region. For the AdS instability problem, the only meaningful calculation is the gravitational self-interaction of a thin-shell when it passes through $r<r_0$. The propagation in the $r>r_0$ region is just propagation in an empty AdS space; the shell simply travels out, reflects off the boundary, and repeats the gravitational evolution near the origin. Since the profile is modified by a small fraction $\sim (\epsilon^2/w)$ during each bounce, we expect on the time scale $\sim \epsilon^{-2}$ an order-one change to accumulate.

For example, the self-interaction might make the shell thinner after each bounce, meaning that the gravitational effect becomes stronger, since more energy is squeezed into a smaller region. If that behavior persists, then eventually the energy will be compressed during a bounce into a region near the origin smaller than its Schwarzschild radius.  At this point, the weak-gravity approximation will break down, and it is very likely that in the $\epsilon^{-2}$ time scale, the shell will evolve into a black hole. On the other hand, it is also possible that the shell becomes wider after each bounce, and energy is dispersed into a larger region. In this case, there is no particular reason why gravitational effects would necessarily become strong and no indication that a black hole would form in the $\epsilon^{-2}$ time scale. The main goal here is to set up a calculation that can capture these two different behaviors.

Before moving on, we need to address the applicability of the thin-shell approximation.  A full dynamical picture should accommodate energy distributions of all thicknesses. However, when $w\sim \rads$, there is no clean way to separate the self-interaction from the effects of the AdS space. Nevertheless, our main interest is the instability in AdS toward black hole formation. In the small-$\epsilon$ limit, the energy must become concentrated into thin shells to even have a chance of eventually forming a black hole. Note that not all of the energy needs to be in one thin shell.  But, the evolution toward a black hole is determined by the shell with the highest radial energy density, which  is dominated by its self-interaction, so we can ignore the influence of other energy distribution outside the shell. 

\subsection{Near-Minkowski expansion}

According to our approximation scheme, we can adopt the weak-gravity expansion in Minkowski space \cite{BizChm08}:
\begin{eqnarray}
\phi &=& \epsilon \phi_0 + \epsilon^3 \phi_1 + ... 
\label{eq-scale}\\
g_{\mu\nu} &=&  g_{\mu\nu}^0 + \epsilon^2 g_{\mu\nu}^1 + ...
\end{eqnarray}
 At zeroth order in $\epsilon$, the background is empty Minkowski space, 
\be
\label{eq-zeroth-order-metric}
\sim\mathcal{O}(\epsilon^0)~, \ \ \ \ 
ds^2 = -dt^2 + dr^2 + r^2d\Omega^2~, 
\ee
into which we put the initial shell profile.  To first order, the equation of motion for $\phi$ is just that of a free field,
\be
\label{eq-first-order-eom}
\sim\mathcal{O}(\epsilon)~, \ \ \ \ 
\ddot\phi_0 - \phi_0'' - \frac{2}{r}\phi_0' =0~.
\ee
At the next order, gravity responds to the stress-energy tensor of the first-order profile.  We therefore must solve the Einstein equation $G_{\mu\nu}=8\pi T_{\mu\nu}$ to leading order in small perturbations around empty Minkowski space. Spherical symmetry excludes dynamical degrees of freedom in the metric, so we only need to solve constraint equations. The $tt$ and $rr$ components suffice to provide the full answer, and the solution is parametrized by two intuitive quantities: enclosed mass $M$ and gravitational potential $V$.
\be
\label{eq-metric1} 
ds^2 = -[1+2\epsilon^2 V(r,t)]dt^2 + \left[1+\frac{2\epsilon^2 M(r,t)}{r}\right]dr^2
+r^2d\Omega^2~.
\ee
Note that we have also explicitly extracted the $\epsilon$ scaling from $M$ and $V$, which are given in terms of the leading-order fields:
\begin{eqnarray}
\sim\mathcal{O}(\epsilon^2)~, \ \ \ \ 
& & \frac{2M'}{r^2} = 8\pi \frac{\dot\phi_0^2+\phi_0'^2}{2}~, \\
& & \frac{2}{r}\left(-\frac{M}{r^2}+V'\right) =
8\pi \frac{\dot\phi_0^2+\phi_0'^2}{2}~,
\end{eqnarray}
with boundary conditions $M(0,t)=0$ and $V(\infty,t)=0$. Finally, the leading nontrivial dynamics comes at the next order---the change in geometry back-reacts on the field profile.
\be
\label{eq-third-order-eom}
\sim\mathcal{O}(\epsilon^3)~, \ \ \ \ 
\ddot{\phi}_1-\phi''_1-\frac{2}{r}\phi'_1
= C\left(\ddot\phi_0+\phi_0''+\frac{2}{r}\phi_0'\right) 
+ \dot{C}\dot\phi_0 + C'\phi_0'~.
\ee
Here we have abbreviated $C=(V-M/r)$.  We see that the field at this order obeys the same wave equation as in the previous order with the addition of a nontrivial source term.

The radial wave equation can be rewritten as a $(1+1)$-dimensional wave equation by introducing $u = r\phi$:
\begin{equation}
r\left(\ddot\phi-\phi''-\frac{2}{r}\phi'\right)
=\ddot{u} - u''~.
\end{equation}
This implies that the initial shell profile given in Eq.~(\ref{eq-init}) is really just the left-moving part of an exact, leading-order solution,
\begin{equation}
r\phi_0(r,t) = u_0(r,t) = \sqrt{w}
\left[f\left(\frac{r-t}{w}\right) - f\left(\frac{-r-t}{w}\right)\right]~.
\label{eq-u0}
\end{equation}
We remind the reader that in Eq.~(\ref{eq-scale}), the $\epsilon$ dependence has been extracted explicitly for $\phi_0$, therefore also for $u_0$. We have taken the liberty to choose the initial time $t_i=-r_0$ to simplify this expression. This allows us to start this calculation once the shell enters the $r<r_0$ region, and the center of the shell reflects off the origin at $t=0$. Later, we will be interested in corrections to the profile at $t_f=r_0$, when the shell is leaving the central Minkowski region. 

Rewriting the system in terms of the (1+1)-dimensional function $u$ is essentially employing a method of images; we extend the range of $r$ into the unphysical $r<0$ region. To implement boundary conditions at $r=0$ such that all physical quantities are finite and smooth, we require $u(r,t)$ to be antisymmetric. Similarly, we can extend the definition of $M$ to negative $r$, 
\begin{equation}
\label{eq-mass}
M(r,t) = \int_0^r d\tilde{r}~ \frac{\dot\phi_0^2+\phi_0'^2}{2}4\pi \tilde{r}^2 ~ ,
\end{equation}
which is naturally an odd function of $r$. The same extrapolation shows that $V$ is an even function of $r$.

In terms of these new variables, the problem of a shrinking shell has been mapped to the problem of two wavepackets colliding at $r=t=0$. Note that this picture is more realistic than it seems; antipodal points of the shell do indeed collide with each other. When the shell is far from the origin, even the leading-order radial energy density is approximately equal to the naive definition of energy in this $(1+1)$-dimensional simplification:\footnote{Note that the total energy $E \equiv M(\infty,t) = \int^\infty_0 \rho(r) dr$ is in fact equal to the naive $(1+1)$-dimensional energy $ \int^\infty_{-\infty} 2\pi \frac{\dot{u}_0^2+u_0'^2}{2}~dr.$}
\begin{equation}
\label{eq-energydensity}
\rho_0= 4\pi r^2 \frac{\dot{\phi}_0^2+\phi_0'^2}{2}~
\approx 4\pi \frac{\dot{u}_0^2+u_0'^2}{2}~.
\end{equation}
To leading order, the colliding shells simply pass through each other. Our goal is to solve the next-order nontrivial effect of such a collision by solving Eq.~(\ref{eq-third-order-eom}), which in terms of $u$ is simply
\begin{equation}
\ddot{u}_1 - u_1'' 
= C\left(\ddot{u}_0+u_0''\right)+\dot{C}\dot{u}_0
+C'\left(u_0'-\frac{u_0}{r}\right)\equiv S(r,t)~.
\label{eq-eomu1}
\end{equation}

This description has a striking resemblance to soliton collisions \cite{ALY13,ALY13a}. The key to this type of problem is that, before solving the equations, we should already anticipate the physical meaning of the answer. At $t_f=r_0$, after the collision, the leading-order solution implies that an out-going shell of the opposite sign reaches exactly $r=r_0$. On top of that, we can organize the next-order correction into the following form:
\begin{equation}
u_0 + \epsilon^2 u_1 = u_0 - \epsilon^2\left( \frac{\partial u_0}{\partial r}\Delta r 
+ \frac{\partial u_0}{\partial w}\Delta w + ... \right)
\label{eq-shellexp}
\end{equation}
We have again extracted the $\epsilon$ dependence explicitly. The shell is actually shifted by $\epsilon^2\Delta r$ from its expected position, its width has changed by $\epsilon^2\Delta w$, and there will be other changes orthogonal to these.

The function $u_1$ at $t_f=r_0$ can be solved from Eq.~(\ref{eq-eomu1}) by integrating the retarded Green's function:
\begin{eqnarray}
\label{u1solution}
u_1(r,r_0) = \frac{1}{2}\int_{-r_0}^{r_0}dt
\int_{r-r_0+t}^{r+r_0-t} dr'~S(r',t)~.
\end{eqnarray}
Note that the lower limit of $r'$ can be negative, which is allowed due to our method of images. The result, however, is the same if we replace the lower bound of the integration range by its absolute value.  

Note that this $u_1$ is only the difference between the incoming shell at $t=-r_0$ and the out-going shell at $t=r_0$, both at position $r=r_0$. Nevertheless, as we have argued that the propagation further to $r=\infty$, the reflection, and the propagation back to $r=r_0$, can all be taken as trivial. This allows us to directly relate $u_1$ to the functional $A$ from Eq.~(\ref{eq-onebounce}), which gives the leading-order change due to one bounce.
\be
\label{Adef}
A\left[\frac{u_0}{r},\frac{\dot{u}_0}{r}\right] 
= -\frac{u_1(\tilde{r},t)}{\tilde{r}}~,
\ee
where $\tilde{r}=2r_0-r$ is the spatial reflection of $r$ around $r_{0}$. The extra minus sign and changing to this ``flipped'' position are due to the trivial propagation to and from $r=\infty$.

The full procedure to calculate $u_1$ and extract physical information like $\Delta r$ and $\Delta w$ are tedious but straightforward. We will present the analytical and numerical details in Appendices \ref{sec-shift} and \ref{sec-examples}. 
Here we highlight two relevant features of the results:

\begin{enumerate}
\item $u_1$ has a $\sim\log r_0$ contribution, which comes entirely from the position shift,
\begin{equation}
\Delta r = -\frac{\int u_1\partial_ru_0~dr}{\int (\partial_ru_0)^2~dr}~,
\end{equation}
which has a clear physical meaning. The leading-order profile $u_0$ follows the $t=|r|$ trajectory, but the next-order correction to the metric modifies the null geodesics. The shell will therefore return to $r=r_0$ not exactly when $t=r_0$. However, this shift is irrelevant to the pertinent question of whether energy gets focused.\footnote{This position shift is related to a shift in frequency in the momentum space analysis observed in other papers \cite{BalBuc14}. 
}
\item The change in the shell's width is given by
\begin{equation}
\label{eq-deltaw}
\Delta w = \frac{\int u_1\partial_wu_0~dr}{\int (\partial_wu_0)^2~dr}~.
\end{equation}
Since we have already scaled out the $\epsilon$ dependence, $\Delta w$ only depends on the shape of the shell (i.e. the function $f$ one chooses in Eq.~(\ref{eq-u0})), and it is independent of both $\epsilon$ and $w$.
\end{enumerate}

In particular, our main result is that $\Delta w$ is just as likely to be positive as negative. Specifically, when we flip the profile of the incoming shell, $f(x) \rightarrow f(-x)$, then $\Delta w\rightarrow-\Delta w$. As a special case, a symmetric profile with $f(x)=f(-x)$ will result in no first order $\Delta w$ during one bounce. This demonstrates that the gravitational self-interaction in AdS is not biased toward focusing energy, and the collapse of small perturbations into black holes is probably not the generic behavior, at least not on time scales $\lesssim\epsilon^{-2}$.  

As a complementary calculation, we also investigate how the maximum radial energy density $\rho_ {\rm Max}$ of the shell behaves under the same $f(x) \rightarrow f(-x)$ transformation.  Like the width $w$, we find that if for a given profile $\rho_{\rm Max}$ increases with each bounce, then for the flipped profile it decreases. This provides another indication that the weak-gravity dynamics are biased neither toward nor against focusing energy.

In the next section, we will give general proofs of these statements. We will also present numerical examples in Appendix \ref{sec-examples}.


\section{Focusing and defocusing}
\label{sec-proof}

\subsection{Antisymmetry of the field correction}
\label{sec-u1flip}

As a preliminary step in proving the statements of the previous section, we need to determine how the first-order correction $u_{1}$ responds to a spatial flip of the initial profile $u_{0}$ such that $f(x) \rightarrow \tilde{f}(x) = f(-x)$. We find that 
\begin{equation}
\label{eq-uoneflip}
u_{1}(r,r_{0}) \rightarrow  \tilde{u}_{1}(r,r_{0}) \simeq - u_{1}(\tilde{r} 
,r_{0}),
\end{equation}
where $\tilde{r}=2r_{0} - r$ is again the spatial reflection of $r$ around $r_{0}$.  Note that this is an approximate statement; for a shell of width $w$, the error in Eq.~(\ref{eq-uoneflip}) is of order $w^2/r_0$. As we argued in Sec.~\ref{sec-weakG}, in the $\epsilon\rightarrow0$ limit, we can choose $r_0$ to make this error arbitrarily small. 

The quantities that enter the expression (\ref{eq-intu1}) for $u_{1}$ are $u_{0}$ and its derivatives and $C$ and its derivatives. So, let us first see how these quantities transform under the flip. From Eq.~(\ref{eq-u0}), we can see that:
\be
\label{eq-u0flip}
u_{0}(r,t) \rightarrow 
\tilde{u}_{0}{(r,t)} = -u_{0}(r,-t).
 \ee
Then, simply by differentiating the two sides of the equation (either with respect to $r$ or $t$), we obtain the same transformation behavior for the derivatives of $u_{0}$.  Now, to see how $C$ transforms, all we need is to determine the transformation of $M$, defined in Eq.~(\ref{eq-mass}):
\begin{eqnarray}
M(r,t) & \rightarrow & \tilde{M}(r,t)  =  2 \pi \int _{0}^{r} dr' \left( \dot{\tilde{u_{0}}} ^{2}(r',t) + \tilde{ u_{0} }^{'2}(r',t) + \left( \frac{ \tilde{u_{0}}(r',t) }{r'} \right) ^{2} - 2\frac{ \tilde{u_{0}}(r',t) \tilde{u_{0}}^{\prime}(r',t) }{r} \right) \nonumber \\
& = & 2 \pi  \int _{0}^{r} dr' \left( \dot{u_{0}}^{2}(r,-t) + u_{0}^{'2}(r,-t) + \left( \frac{u_{0}(r,-t)}{r'} \right) ^{2} - 2\frac{u_{0}(r',-t)u_{0}^{\prime}(r,-t)}{r'} \right)  \nonumber \\
& = & M(r, -t). 
\end{eqnarray}
Since $V$ has the same behavior as $M$, then $C(r,t) = V - \frac{M}{r}$ transforms under the flip as:
\begin{equation}
C(r,t) \rightarrow \tilde{C}(r,t) = C(r,-t).
\end{equation}
Again, a similar relation holds for the derivatives of $C$. Combining the above results, we see that the source term $S(r,t)$, defined in Eq.~(\ref{eq-eomu1}), behaves as
\begin{equation}
S(r,t) \rightarrow \tilde{S}(r,t) = - S(r,-t)
\end{equation}
under flipping of the initial profile. Also, by demanding regularity at the origin $r=0$, the initial profile is antisymmetric in $r$, which in turn implies that $M(r,t)$ is also antisymmetric in $r$; hence $C(r,t)$ is symmetric. These properties imply the antisymmetry of $S(r,t)$ in its first argument, $S(r,t) = - S(-r,t)$.

\begin{figure}[tb]
\begin{center}
\includegraphics[width=14cm]{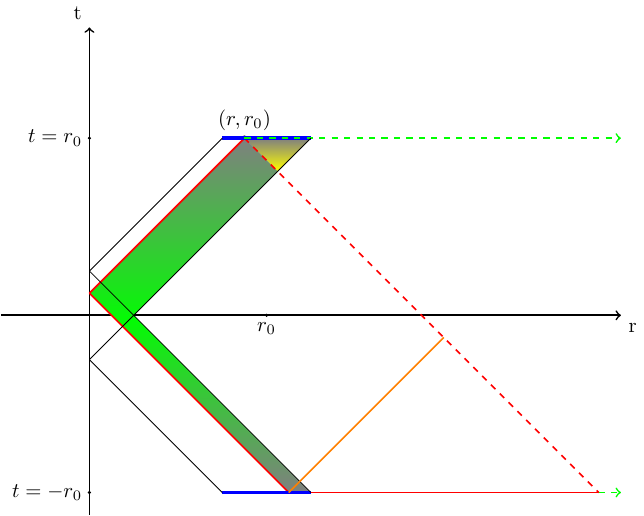}
\caption{The areas of integration for the computation of $u_1(r,r_{0})$. The blue solid line represents the source at times $t= \pm r_{0}$. The red lines (solid and dashed) indicate the actual area of integration in Eq.~(\ref{eq-intu1}) ({\it i.e.} the integral $ \int_{-r_{0}}^{r_{0}} dt \int_{\left| r-r_{0}+t \right|}^{r+r_{0} -t} d \tilde{r}$), and the green shaded region indicates where the integrand is nonzero. The solid red lines together with the dashed green lines correspond to the region of integration $\int_{-r_{0}}^{r_{0}} dt \int_{\left| r-r_{0}+t \right|}^{\infty} d\tilde{r}$ used in our approximate Eq.~(\ref{eq-approximateu1}).  The yellow triangle at the top shows the extra nonzero contribution included in the second integral, which is suppressed by $\frac{w^2}{r_{0}}$. Alternatively, the integral can be approximated by using the orange line instead of the horizontal red line.
\label{fig:spacetimediagram}}
\end{center}
\end{figure}

Now we are ready to prove Eq.~(\ref{eq-uoneflip}), starting from the integral expression Eq.~(\ref{eq-intu1}) for $u_1$.  The integration regions are illustrated in Fig.~\ref{fig:spacetimediagram}. 

 We first make an approximation to Eq.~(\ref{eq-intu1}).  The upper limit of the $r'$ integral is $r+r_{0} -t$. Instead, we will extend the region of integration up to $r' = \infty$. Because the wavepacket has compact support only over a region of width $w$, the error introduced by this approximation comes just from the yellow shaded triangle in Fig.~\ref{fig:spacetimediagram}. The area of this added triangle is $\mathcal O(w^2)$ and, since $C(r,t) \sim \frac{1}{r}$, the integrand is of order $\frac{1}{r_{0}}$.  Hence, the error is suppressed by a factor of $\frac{w^2}{r_{0}}$.  

A similar, and perhaps even more physical, approximation, albeit with more cumbersome limits of integration, can be made by considering the area of integration denoted by the red lines together with the orange line in Fig.~\ref{fig:spacetimediagram}. In that case, instead of adding the extra contribution from the yellow triangle at the top, we would subtract the area of the green triangle at the bottom. However, the results would be the same. 

After this approximation, we have:
\begin{equation}\label{eq-approximateu1}
u_{1}(r,r_{0}) \simeq  \frac{1}{2}\int_{-r_{0}}^{r_{0}} dt \int_{\left|r-r_{0} + t \right| }^{\infty} dr'~{S}(r',t)~. 
\end{equation}
Now, flipping the initial profile we get:
 \begin{eqnarray}
 \tilde{u}_{1}(r,r_{0})  & \simeq &   \frac{1}{2}\int_{-r_{0}}^{r_{0}} dt \int_{\left|r-r_{0} + t \right| }^{\infty} dr'~ \tilde{S}(r',t)~.
 \end{eqnarray}
Using the flipping property of $S(r,t)$, as discussed above, we can write:
\begin{eqnarray}                 
\tilde{u}_{1}(r,r_{0})  &\simeq &  - \frac{1}{2}\int_{-r_{0}}^{r_{0}} dt \int_{\left|r-r_{0} + t \right| }^{\infty} dr'~ S(r',-t)~.
\end{eqnarray}
We can now change the dummy integration variable $t$ to $-t$ and use the relation $r = 2r_{0} - \tilde{r}$ to rewrite the lower integration limit, giving
\begin{eqnarray}
 \tilde{u}_{1}(r,r_{0})  &\simeq &  -\frac{1}{2}\int_{-r_{0}}^{r_{0}} dt \int_{\left|(2r_{0} - \tilde{r})-r_{0} + t \right| }^{\infty} dr'~S(r',t)~.
\end{eqnarray}
Comparing this expression to Eq.~(\ref{eq-approximateu1}), we obtain
 \be
\tilde{u}_{1}(r,r_{0}) = - u_{1}(\tilde{r},r_{0})~. 
 \ee
 which is our desired result.

\subsection{Shell width}

In this subsection we will prove Eq.~(\ref{eq-deltaw}); that is, under a spatial flip $f(x)\rightarrow \tilde{f}(x) =  f(-x)$, the leading-order correction to the width is antisymmetric:
\be
\label{Deltawflip}
\Delta w \rightarrow \Delta \tilde{w} = -\Delta w
\ee
We assume the profile $u_0$ has compact support within $r_0-w/2<r<r_0+w/2$, and evaluate $\Delta w$ at late time $t_{f} = r_{0}$, well after the collision, at which point the left-moving and the right-moving wavepackets are far away from $r = 0 $ and do not interfere with each other. In that case, when computing $\Delta w$ from Eq.~(\ref{eq-deltaw}), we can just integrate over the right-moving wavepacket; integrating over both wavepackets would just double both the the numerator and denominator in Eq.~(\ref{eq-deltaw}), yielding the same result.  The expression for the change in width of the flipped profile is then
 \be
 \label{Deltaw}
\Delta \tilde w = \frac{ \int_{r_{0}-w/2}^{r_{0} + w/2} dr~\tilde u_{1}(r, r_{0})\partial _{w}\tilde u_{0}(r, r_{0})}{\int_{r_{0}-w/2}^{r_{0} + w/2} dr~(\partial_w \tilde u_0)^2} \ .
\ee

It is convenient to define $y = r-r_0$, such that the spatial flip of the initial profile is given by 
\be
u_{0}(r_{0} + y, r_{0}) \rightarrow \tilde{u}_{0}(r_{0} + y, r_{0}) = u_{0}(r_0-y,r_{0}) ~ .
\ee
Starting with the numerator of Eq.~(\ref{Deltaw}) and using the properties of $u_0$ and $u_1$ under the flip, we can write
  \begin{eqnarray}
  \int_{r_{0}-w/2}^{r_{0} + w/2} dr~ \tilde{u}_{1}(r, r_{0}) \partial _{w}\tilde{u}_{0}(r, r_{0})  
  &=& \int_{-w/2}^{w/2} dy~\tilde{u}_{1}( r_{0} + y, r_{0})\partial _{w}\tilde{u}_{0}(r_{0} + y, r_{0}) \nonumber \\
                         & = & - \int_{-w/2}^{w/2} dy~u_{1}(r_{0} - y, r_{0}) \partial _{w}u_{0}(r_{0} - y, r_{0})  \nonumber \\
                         & = & - \int_{-w/2}^{w/2} dy~u_{1}(r_{0} + y, r_{0}) \partial _{w}u_{0}(r_{0} + y, r_{0})                       
 \end{eqnarray}
In the third line, we changed the dummy integration variable from $y$ to $-y$. The flip therefore changes the sign of the numerator.  Following these same steps with the denominator of Eq.~(\ref{Deltaw}), we can see that it is invariant under the flip.  Putting these two statements together yields the desired result, $\Delta \tilde w = -\Delta w$.

\subsection{Energy density}

A similar argument holds for the leading-order change in the energy density $\Delta \rho$ at time $t_f = r_0$ due one bounce through the origin. Specifically, for $f(x) \rightarrow \tilde{f}(x) = f(-x)$, we find
  \be
  \label{eq-deltarhoflip}
\Delta \tilde{\rho}(r,r_0) \simeq - \Delta \rho (\tilde{r},r_0) ~ .
  \ee
where recall $\tilde{r} = 2r_0 - r$. The full radial energy density far from the origin is approximately the (1+1)-dimensional expression, as in Eq.~(\ref{eq-energydensity}). Expanding it to the next order, we find
 \be
 \rho_0 + \epsilon^2 \Delta \rho  =  4\pi  \frac{ \dot u_0^2 + {u_0'}^2}{2} +4\pi \epsilon^2 \left(  \dot u_0\dot u_1 + u_0' u_1' \right) ~ .
 \ee
We kept our principle of always extracting $\epsilon$ explicitly. The first term is the initial energy density given in Eq.~(\ref{eq-energydensity}), which is actually $\epsilon^{-2}$ times the actual physical energy density. The second term is the leading change due to a single bounce.

The formula (\ref{eq-uoneflip}) we found for the behavior of $u_1$ under the flip holds at the specific time $t=r_{0}$, and it is not straightforward to see that the same relation holds for $\dot u_1$.  An alternative way to proceed is to include the explicit expression for the derivatives of $u_{1}$ at $t_{f} = r_{0}$:
\begin{eqnarray}
u_{1}'(r,r_{0}) & = & \frac{1}{2} \int_{-r_{0}}^{r_{0}} dt~ 
\left[ S(r+r_{0} -t) -S(r-r_{0} + t) \right] \\
\dot{u}_{1}(r,r_{0}) & = & \frac{1}{2} \int_{-r_{0}}^{r_{0}} dt \left[ S(r+r_{0} -t) + S(r-r_{0} + t) \right] ~ . 
\end{eqnarray}
Using these formulae and the expression for $u_{0}$ Eq.~(\ref{eq-u0}), and omitting irrelevant constants, 
 we can write down the explicit expression for $\Delta \rho$.  
\begin{eqnarray}
\Delta \rho(r,r_{0}) 
                                    & = &  \int_{-r_{0}}^{r_{0}} dt~\left[ f^{'}(-r-r_{0}) S(r+r_{0} -t,t) - f^{'}(r-r_{0}) S(r-r_{0} +t,t) \right]
\end{eqnarray}
Since the function $f(x)$ has compact support of width $w$ around $x=0$ and we consider values of $r$ on the order of $r_{0}$, then the first term in the integrand vanishes.
%
%
We can now determine how $\Delta \rho$ behaves under $f(x) \to f(-x)$:
\be
\Delta \tilde{\rho}(r,r_{0}) = - \int_{-r_{0}}^{r_{0}} dt  f^{'}(-r+r_{0}) S(r-r_{0} +t,-t) ~ . 
\ee
Changing the dummy integration variable $t$ to $-t$ and
%
substituting with $r=2r_{0} - \tilde{r}$, we obtain
\be
\Delta \tilde{\rho}(r,r_{0}) =  - \int_{-r_{0}}^{r_{0}} dt f^{'}(\tilde{r}-r_{0}) S(-\tilde{r}+r_{0} -t,t) . \nonumber
\ee
From the antisymmetry of $S$ in its first argument, we get
\begin{eqnarray}
\Delta \tilde{\rho}(r,r_{0}) & = &  + \int_{-r_{0}}^{r_{0}} dt   f^{'}(\tilde{r}-r_{0}) S(\tilde{r}-r_{0} +t,t)  \nonumber \\
                                        & = & - \Delta \rho (\tilde{r},t),
\end{eqnarray}
which is indeed Eq.~(\ref{eq-deltarhoflip}).

Eq.~(\ref{eq-deltarhoflip}) relates the change in energy density at an arbitrary point $r$ and its image $\tilde{r}$ under the flip. However, we are particularly interested in how the change in the maximum energy density is affected by the flip.

%
The energy density at the position of the maximum, $r_ {\rm Max}$, after one bounce, can be expanded as
%
\be
 \rho_ {\rm Max} \equiv \rho(r_ {\rm Max}) =  \rho^{(0)}(r_ {\rm Max}) + \epsilon^2 \Delta \rho(r_ {\rm Max}) \ .
\ee 
It might be tempting to directly identify this with the change of maximum energy density. However, we should remember that in addition, the location of the maximum $r_ {\rm Max}$ is also, in general, affected by the bounce, receiving corrections at the same order, $r_ {\rm Max}  = r^{(0)}_ {\rm Max} + \epsilon^2 \Delta r_ {\rm Max}$.  So, expanding to order $\epsilon^2$, we find
\be
\rho_ {\rm Max}^{(0)} + \epsilon^2 \Delta\rho_ {\rm Max}  = \rho^{(0)}\left(r^{(0)}_ {\rm Max}\right)+ \epsilon^2 {\rho^{(0)}}'\left(r^{(0)}_ {\rm Max}\right) \Delta r_ {\rm Max}  + \epsilon^2 \Delta \rho\left(r^{(0)}_ {\rm Max}\right) \ .  
\ee
However, $r_ {\rm Max}$ is an extremum of $\rho^{(0)}$, and so ${\rho^{(0)}}'\left(r^{(0)}_ {\rm Max}\right) = 0$.  To leading-order, the change in the maximum is due only to a change in $\rho$ and not a shift in the location of the maximum:
\be
\Delta\rho_ {\rm Max} = \Delta \rho\left(r^{(0)}_ {\rm Max}\right) \ .
\ee

Now, under a flip, $r^{(0)}_ {\rm Max}$ is mapped to ${\tilde{r}^{(0)}}_ {\rm Max} = 2r_0 - r^{(0)}_ {\rm Max}$, the location of the maximum of $\tilde\rho$; that is,
\be
\Delta\tilde\rho_ {\rm Max} = \Delta\tilde\rho\left({\tilde{r}^{(0)}}_ {\rm Max}\right) \ .
\ee
From Eq.~(\ref{eq-deltarhoflip}), we can see that 
\be
\Delta\tilde\rho_ {\rm Max} = - \Delta\rho\left(r^{(0)}_ {\rm Max}\right) =  - \Delta\rho_ {\rm Max} \ .
\ee
Therefore, flipping the profile reverses the direction of the change in the maximum energy density.  This result nicely complements our result regarding the the change in width Eq.~(\ref{Deltawflip}).  Both of these results indicate that there is no bias in the weak-gravity dynamics toward either increasing or decreasing the energy concentration.



\section{Discussion}
\label{sec-dis}

\subsection{Phase space diagram}

In Sec.~\ref{sec-weakG}, we provided the recipe to compute the change in a field profile after one bounce, and it contained all the information about the functional $A$ in Eq.~(\ref{eq-onebounce}). In principle, one can add the resulting $\bar{\phi}_1$ to the original $\bar{\phi}_0$ to make a new initial condition, and calculate the result of the next bounce. Choosing a small $\epsilon$ and reiterating this process $\sim\epsilon^{-2}$ times is equivalent to solving Eq.~(\ref{eq-contscale}). In principal, this will directly reproduce the long term evolution. Unfortunately, there is one technical difficulty that we have not been able to overcome.

Our method has one disadvantage: energy conservation is by definition an approximation. We basically ``turn on'' a self-gravitational potential when a shell shrinks below $r_0$, let energy flow between it and the field kinetic terms, then turn it off when the shell expands over $r_0$. The amount of potential energy we turn on and off differs by $\sim \epsilon^4w/r_0^2$. Although this is suppressed by an extra factor of $w/r_0$ from the quantities we care about in Sec.~\ref{sec-proof} and so it does not invalidate our results, it is technically difficult to control. We could na\"ively go to a larger $r_0$ for better energy conservation, which would increase the integration range required to solve Eq.~(\ref{eq-eomu1}).  However, more numerical resources would then be necessary in order to proceed.

\begin{figure}[tb]
\begin{center}
\includegraphics[width=0.7\textwidth]{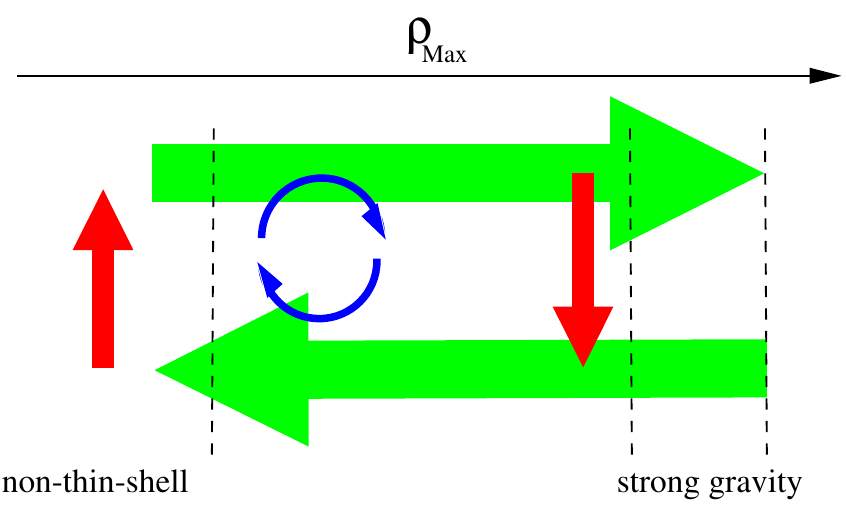}
\caption{A two-dimensional projection within a constant-energy slice of the phase space. The horizontal axis is the peak energy density, and the big, green arrows toward left and right represent the focusing and defocusing flow due to gravitational self-interaction. Together with the upward and downward flows represented by the small, red arrows, the phase space has a circular flow pattern. The blue loop represents quasi-periodic solutions that stay within the center of this circular flow.
\label{fig-flow}}
\end{center}
\end{figure}

Nevertheless, we might have learned enough about what happens in a single bounce to make a reliable extrapolation. We will attempt to do so by drawing a phase-space diagram. Since there are no gravitational degrees of freedom within spherical symmetry, the phase space of perturbations is given by all possible scalar field profiles. Due to energy conservation, we can focus on one fixed-energy, co-dimension-one surface in this infinite dimensional phase space. Within this surface, we can draw a two-dimensional projection (Fig.~\ref{fig-flow}) and understand its structure based on our knowledge of the dynamics of one bounce.

One guiding principle of this diagram is that during one bounce, the profile changes by an infinitesimal amount $\sim\epsilon^2$, which is also an infinitesimal distance in the diagram. Within the weak-gravity time scale $\sim\epsilon^{-2}$, the evolution trajectory covers a finite distance of the diagram. In this way, the diagram directly represents the dynamical evolution in the rescaled time as given by Eq.~(\ref{eq-contscale}).

The horizontal axis of this two-dimensional diagram represents ``how close is this profile to becoming a black hole''. More technically, it is quantified by the maximum radial energy density at $r=0$ that is reached during one AdS time. In the small-$\epsilon$ limit, the profile is basically freely propagating, so this is a well-defined quantity. Heuristically, this maximum is reached when the highest ``peak'' goes through $r=0$, and its value depends on the height of this peak, $\rho_{\rm Max}$. 

Note that throughout this paper, we have been referring to $\rho$ as the rescaled energy density. In our conventions, the actual physical energy density is given by $\epsilon^2\rho$. It is still convenient to consider the rescaled density here, since $\rho_{\rm Max}$ quantifies how much higher this peak is than the average, namely the relative concentration of energy. Its value increases toward the right-hand-side of Fig.~\ref{fig-flow}. For any finite $\epsilon$, there is a finite value of $\rho_{\rm Max}\sim\epsilon^{-2}$ that represents a density high enough to become a black hole. It can be drawn as a vertical line. Some finite distance to the left of this line, we have another line signifying that the energy density is high enough to make gravity too strong to be described by the weak-gravity  expansion. 

To the left of this second line, gravity is weak enough that our analysis applies. Somewhere even further to the left, our approximation starts to fail for a different reason:  we can no longer describe this peak as an isolated thin shell satisfying the hierarchy in Eq.~(\ref{eq-hierarchy}). In the small-$\epsilon$ limit, the region to apply our method always exists. This left boundary is not a very clear line. Nevertheless, in this diagram we can roughly picture it as $\rho_{\rm Max}R_{\rm AdS}\sim1$, that the maximum peak density is comparable to the average density. Clearly, energy is too evenly distributed in the entire AdS space that nothing could be treated as an isolated thin shell.

The vertical axis of this diagram is not intended to represent any particular parameter of the field profile. It is merely reflecting the fact we established in Sec.~\ref{sec-proof} that focusing and defocusing dynamics are equally likely in one bounce. This means one can always find some parameter such that the middle region is divided into two halves: in the upper half, the evolution makes the peak grow higher and moves closer to forming a black hole, and in the lower half, the peak gets lower and moves away from forming a black hole. In App.~\ref{sec-examples}, we give specific numerical examples and argue the parameter controlling focusing and defocusing is the asymmetry of energy distribution: focusing occurs when the shell is is denser in the leading edge, and defocusing when it is denser in the tail.

In addition to focusing and defocusing which correspond to flowing horizontally in Fig.~\ref{fig-flow}, what  tendencies to flow in the vertical direction can we identify? Generally speaking, when $\rho_{\rm Max}$ is large, on the left side of Fig.~\ref{fig-flow}, the system will tend to flow upwards. This is because a shrinking peak cannot remain the highest peak forever: a growing peak with smaller initial height will eventually take over. If that were the only vertical motion, it would lead to only two possible trajectories: starting in the upper half, $\rho_{\rm Max}$ would keep increasing and flow directly toward black hole collapse; alternatively, staring in the lower region, $\rho_{\rm Max}$ would first decrease, then bounces back to become a black hole. This directly violates many numerical results, so there must be a downward flow somewhere in Fig.~\ref{fig-flow}.

\begin{figure}[tb]
\begin{center}
\includegraphics[width=0.7\textwidth]{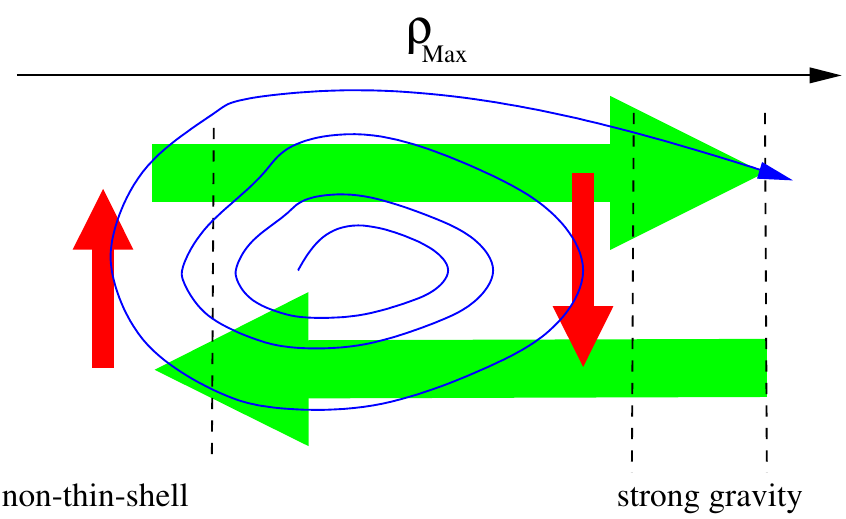}
\caption{The same circular flow in the phase space, but the blue trajectory now represents an unstable solution. Though initially it follows the circular flow, it fluctuates to larger radius, eventually enters the strong gravity region, and collapses into a black hole.
\label{fig-unstable}}
\end{center}
\end{figure}

The two numerical examples in App.~\ref{sec-examples} provide tentative evidence for a downward flow. What we see is that given a symmetric profile on the boundary between the upper and lower region, after one bounce it picks up an asymmetry similar to the profiles in the lower region---its energy becomes denser in the tail. Of course, we studied only two one-parameter slices through an infinite dimensional phase space, so better numerical and/or analytical investigation is required to verify this. At the level of this paper, we will simply conjecture that such a downward flow exists, because the resulting circulatory flow, shown in Fig.~\ref{fig-flow}, explains existing numerical results very well:
\begin{itemize}
\item The quasi-periodic solutions stay within the circular flow near the center, as in Fig.~\ref{fig-flow}.
\item The unstable solutions initially stay within the circular flow, but their radii vary wildly and eventually these solutions enter the strong gravity regime, as in Fig.~\ref{fig-unstable}
\end{itemize}

Note that the actual motion in the true phase space is still very complicated. In this is a two-dimensional projection, evolution trajectories are allowed to cross each other. Nevertheless, this circular flow allows us to better visualize the dynamics in the phase space. 


We can also repeat the argument in Sec.~\ref{sec-review} in a more pictorial manner. As one reduces $\epsilon$, most of this diagram does not change. Due to the scaling behavior, all trajectories to the left of the strong gravity line remain the same, and so most of the stable solutions remain stable. The trajectories for unstable solutions must cross the strong gravity line to form black holes, so they potentially can change. 

Actually, the location of the strong gravity line shifts to the right when $\epsilon$ decreases. As the total energy is reduced, it needs to be increasingly focused in order for gravity to become strong, and a collapsing solution must therefore to evolve further, across the new weak-gravity regime. In the $\epsilon\rightarrow0$ limit, black hole formation is equivalent to a weak-gravity evolution in which $\rho_{\rm Max}$ goes all the way to infinity. 

Although the trajectory for a black hole collapse appears divergent, it does not mean that we can immediately rule out such an evolution.  In fact, this divergence is an artifact of the parameter choice, and we should appreciate that  $\rho_{\rm Max}=\infty$ is not an infinite distance away in phase space. 
Recall that $\Delta w$ in one bounce is independent of $w$, so the change in width need not be a small fraction of the total. It is certainly possible to have a profile such that after $\epsilon^{-2}$ bounces, $\Delta w$ is negative and order one, leading to a diverging $\rho_{\rm Max}$.

The real advantage of this picture is that it recasts the $\epsilon\rightarrow0$ limit of the stability problem into the global regularity problem of determining whether $\rho_{\rm Max}$ diverges.  Interestingly, analyzing the regularity of $AdS_3$ perturbations at finite $\epsilon$ below the black hole mass gap is similarly a question of determining whether the energy density diverges. In that case, there is already strong evidence to support regularity\cite{BizJal13,Jal13}. One might hope to reproduce this $AdS_3$ result in higher dimensions in order to confirm that the instability corners indeed shrink to measure zero. We should again caution that spectral analysis can only provide necessary conditions; it can rule out an instability, but it cannot provide equally strong evidence to support one. If the power spectrum of perturbations agrees with a diverging $\rho_{\rm Max}$, a long-time evolution of Eq.~(\ref{eq-diffscale}) in position space is still required for the final answer to the AdS stability problem.

\subsection{Holographic thermalization}

One motivation for studying the stability properties of AdS is to try to learn something about the non-equilibrium dynamics of closed systems. This is due to the AdS/CFT correspondence, which relates this classical gravitational system to the dynamics of a strongly-coupled quantum system.  Most investigations of holographic thermalization study the Poincar\'e patch of AdS, which has an infinite boundary (see, for example, \cite{LinShu08,BhaMin09,AlbJoh10,BalBer11}).  In these cases, any nonzero energy density in the bulk will collapse into a black hole, corresponding to thermalization on the boundary.

Here, instead, we are considering global AdS which has a closed, spherical boundary and therefore a very different thermalization behavior. Other studies of global AdS, such as \cite{AbadaS14}, implicitly assume the connection between forming a black hole in the bulk and thermalization of the boundary system. Although that is valid in some cases, we would like to highlight other possibilities.  What are the possible holographic dual descriptions of the bulk story presented here?

One caveat is that explicit examples of the AdS/CFT correspondence usually contain compact extra dimensions whose sizes are comparable to $\rads$, for example in $AdS_5 \times S_5$. In the $\epsilon\rightarrow0$ limit, the five-dimensional AdS-Schwarzschild black hole has a horizon radius much smaller than $\rads$ and is therefore too small to represent the most typical states; a ten-dimensional black hole of the same total energy, which breaks the symmetry of the $S^5$, has even higher entropy. Therefore, the gravitational stability problem of $AdS_5$ does not directly translate to the thermalization problem in the boundary system. This might be an interesting direction for future work, but we will set this concern aside for now. Let us take a very optimistic point of view that the AdS/CFT correspondence can work with extra dimensions arbitrarily smaller than $\rads$, or even without them.

After limiting our attention to the AdS space and treating our classical field theory as a limit of a quantum gravity theory, the $\epsilon\rightarrow0$ limit leads to a different issue. Recall the well known Hawking-Page transition\cite{HawPag82}: A black hole does not always dominate the micro-cannonical ensemble; given low enough energy, thermal gas is the most typical state. In this case, forming a black hole does not imply thermalization. This is the main issue we wish to clarify.

First of all, this issue highlights the importance of our position space approach. Focusing on the power spectrum, initial conditions occupying only low frequency modes must propagate to higher frequency in order to approach either a black hole or a thermal gas state. This type of turbulent cascade is a necessary condition for thermalization. However, without differentiating between the final states toward which the system could be evolving, one cannot argue unequivocally for or against thermalization.

Next, let us analyze under what circumstances the black hole or the thermal gas state will dominate the ensemble.  For simplicity, we will work via dimensional analysis and ignore any order-one factors. First, note that the $\epsilon\rightarrow0$ limit is actually the weak-gravity limit, corresponding to
\be
\beta\equiv \frac{\rads}{\rs} \gg 1~.
\label{eq-weakG}
\ee
Namely, the Schwarzschild radius of the black hole made by collapsing the scalar field energy is much smaller than the AdS size. On the other hand, the most straightforward standard for trusting classical gravity is
\begin{equation}
\gamma \equiv \frac{\rs}{\lp} \gg 1
\label{eq-classical}
\end{equation}
where we have restored the Planck scale $l_{\rm Planck}=\sqrt{G\hbar}$, which has been set to one in the rest this paper. This condition implies that, at the very least, if a black hole forms, it could be described by classical gravity.  For the limit we have been considering, both $\beta$ and $\gamma$ have been taken to infinity.  We will see that whether the black hole or the thermal gas dominates depends on the details of how that limit is taken.

The entropy of a black hole with energy $E$ is\footnote{Note that we are assuming here that the spacetime is effectively $AdS_4$ at distance scales of order the size of the black hole; in string constructions, such as $AdS_4 \times S_7$, black holes whose radius is small compared to the AdS radius would be eleven-dimensional rather than four-dimensional, leading to different formulas.}
\be
\sbh \sim \left( \lp E \right)^2 \sim \left( \rs \over \lp \right)^2~.
\ee
The entropy of a thermal gas in AdS space with the same total energy is
\begin{equation}
S_{\rm gas} \sim \left(\frac{\rs  \rads}{\lp^2}\right)^{3/4}~.
\end{equation}
Thus, black hole states dominate the micro-canonical ensemble when 
\begin{equation}
\left(\frac{\rs}{\lp}\right) > \left(\frac{\rads}{\rs}\right)^{3/2}~; \ \ \ \gamma > \beta^{3/2}~.
\label{eq-BHd}
\end{equation}
This condition is equivalent to comparing the thermal wavelength $\lambda_{\rm th}$ of the gas to the black hole radius; the black hole dominates the ensemble if
\be
\rs > \lambda_{\rm th} ~.
\ee
We can see that whether the condition in Eq.~(\ref{eq-BHd}) is satisfied depends on how the limits of large $\beta$ and large $\gamma$ are taken. A classical and small-$\epsilon$ limit does not restrict the system to being dominated by either the thermal gas or black hole states.

Note that whether, and for how long, classical evolution is a good approximation depends on more details of the state. For example, even if a black hole forms which is classical according to Eq.~(\ref{eq-classical}), the process by which it formed might not be. A simple rule of thumb for the validity of the classical limit is that the occupation numbers in the modes of interest have to be large. If the system is in a state where the energy is roughly equipartitioned between a number of modes up to some maximum frequency $\omega_{\rm max}$, we require
\be
{\rm energy \ per \ mode} \gg \omega_{\rm max} \label{eq-wmax}~.
\ee
The thermal gas states can never satisfy this condition because modes with frequency of order the temperature have occupation numbers of order one, yet contribute a significant fraction of the entropy of the gas. Independent of the limiting procedure and which states dominate, the thermal gas final state is never compatible with a classical description.\footnote{Nevertheless, from the position-space viewpoint, classical evolution may still describe the ``process of approaching'' a thermal gas state, at least distinguishing it from approaching a black hole. In the later case energy becomes more concentrated, but in the former case it does not.}

On the other hand, a spherically symmetric collapse into a black hole can often be completely classical. Such a process only needs to excite the radial modes from the longest wavelength $\sim R_{\rm AdS}$ to the shortest wavelength $\sim R_{\rm BH}$, thus
\begin{equation}
{\rm number \ of \ modes} = \frac{R_{\rm AdS}}{R_{\rm BH}}~.
\end{equation}
The condition on occupation numbers, Eq.~(\ref{eq-wmax}), becomes
\begin{equation}
\beta \ll \gamma^2~.
\label{eq-sphclassical}
\end{equation}
Comparing this to Eq.~(\ref{eq-BHd}), we see that Eq.~(\ref{eq-sphclassical}) is always true when the black hole dominates the ensemble, but it can still be true even if thermal gas dominates.\footnote{Note that spherical symmetry is very important here. Without it, the number of modes would have been $\left(\frac{\rads}{\rs}\right)^3$, and with that many modes, the black hole collapse would have failed to remain classical in the thermal gas-dominated regime.} {\it Thus, the specific stability problem within classical gravity investigated in this paper, namely a spherically symmetric collapse into a black hole, is a valid dual to some boundary system, independent of whether such a process is equivalent to a efficient thermalization or not.}\footnote{In this paper, ``efficient'' means that thermalization happens in the shortest time scale allowed by the dynamics, $\sim\epsilon^{-2}$. One should not confuse this with, for example, the much faster thermalization in the Poincar\'e patch of AdS, where, within one AdS time, perturbations cross the horizon, form a planar black hole, and appear to thermalize.}

Furthermore, when the thermal gas dominates, if a black hole really forms in the time scale we investigated,
\begin{equation}
T_{\rm weak \ gravity} = R_{\rm AdS}\frac{R_{\rm AdS}}{R_{\rm BH}} \propto \epsilon^{-2}~,
\end{equation}
it could represent a significant delay to thermalization.  In order to confirm this, we need to compare the na\"ive thermalization time $T_{\rm weak \ gravity}$ to the  black hole lifetime, given by the the evaporation time scale,
\begin{equation}
T_{\rm evaporation} = \frac{R_{\rm BH}^3}{l_{\rm Planck}^2}~.
\end{equation}
When $T_{\rm evaporation}>T_{\rm weak \ gravity}$, which requires
\begin{equation}
\label{eq-prethermalization}
\beta < \gamma~,
\end{equation}
the system thermalizes only after forming a long-lived black hole, which eventually evaporates. This process of thermalization via a quasi-stationary thermal-like state is known as prethermalization and has been observed in finite-sized, isolated quantum systems \cite{TroChe11,GriKuh11}.  Note that Eq.~(\ref{eq-prethermalization}) is compatible with thermal gas domination and a classical collapse.


To summarize, spherically symmetric black hole formation within $T_{\rm weak \ gravity}$ can have two different holographic interpretations:
\begin{itemize}
\item When $\gamma^2 > \beta^3$, it represents efficient thermalization of the boundary system.
\item When $\beta < \gamma < \gamma^2 < \beta^3$, it represents prethermalization, which delays true thermalization (to thermal gas\footnote{Note that since thermal gas is never classical, we do not know exactly when will it really form. We only know that within $T_{\rm evaporation}$, the systems was too busy forming a black hole and then remaining as one, so it cannot reach the thermal gas state yet.}) at a time scale $\gtrsim T_{\rm evaporation}>T_{\rm weak \ gravity}$.
\end{itemize}
For the remaining possibility, when $\gamma^2 < \beta^3$ but $\beta > \gamma$, the implication of black hole formation is inconclusive from a thermalization point of view. Black holes decay too fast to be  quasi-stationary intermediate states, but their evaporation cannot guarantee reaching the thermal gas state either. 

\section{Summary}
\label{sec-sum}

\begin{itemize}
\item By combining existing numerical data with our analysis, we have argued that for a massless scalar field in AdS space, in the small-amplitude $\epsilon\rightarrow0$ limit, solutions remaining stable up to the interaction time scale $T\sim\epsilon^{-2}$ form an open set. This improves similar observations in finite-$\epsilon$ numerical simulations \cite{BucLie13,MalRos13} and argues against the conjecture that the weakly turbulent instability occurs in all but a set of measure zero in the space of initial conditions \cite{BizRos11,DiaHor11,DiaHor12}.
\item One important difference between our approach and previous work is that we analyzed the problem in position space. We pointed out that only position space properties can provide necessary and sufficient conditions for the collapse into a black hole. Any analysis of the power spectrum can at most provide necessary conditions for black hole formation.
\item In the position space analysis, we exploited the small-amplitude $\epsilon\rightarrow0$ limit and argued in Sec.\ref{sec-weakG} that the only relevant dynamics are the gravitational self-interactions near $r=0$. This argument requires a hierarchy of scales given in Eq.~(\ref{eq-hierarchy}), which is difficult to reach in realistic numerical simulations. 
\item We showed that gravitational self-interaction near $r=0$ obeys an exact antisymmetry under time reversal. As a result, it is equally likely for interactions to focus or defocus energy. This equality is consistent with existing numerical results.\footnote{More specifically, one could take any numerical simulation and pause it at a moment when gravity is still weak. If one keeps the field profile but reverses the time derivative at this moment, the simulation will literally evolve backward toward the original initial profile, up to the numerical error and higher-order effects (which are small if the hierarchy of scales in Eq.~(\ref{eq-hierarchy}) is satisfied during the process).} We remind the readers that gravity can be effectively repulsive: tidal forces tend to pull things apart. The possible defocusing of a radiation shell is due to such tidal effect. 
\item By making use of scaling symmetry, we simplified the stability problem in the $\epsilon\rightarrow0$ limit into a global regularity problem within a finite rescaled time.  The evolution was recast as a simple, first-order differential equation. We hope that this point of view, combined with the other techniques in this work and the existing literature, will allow a rigorous analysis of stability in the vanishing amplitude limit.
\item Even if black holes do form in the $\epsilon^{-2}$ time scale, we point out that it does not always represent efficient thermalization of the boundary theory via AdS/CFT duality. In some cases, black hole formation describes prethermalization, and actual thermalization is delayed until this black hole evaporates.
\end{itemize}


\acknowledgments

We thank Jan de Boer, Avery Broderick, Alex Buchel, Stephen Green, Diego Hofman, Matthew Johnson, Luis Lehner, Vladimir Rosenhaus, Andrzej Rostworowski, Jorge Santos and Ferdinand Verhulst for useful discussions. This work is part of the $\Delta$-ITP consortium and also supported in part by the Foundation for Fundamental Research on Matter (FOM), both are parts of the Netherlands Organisation for Scientific Research (NWO) that is funded by the Dutch Ministry of Education, Culture and Science (OCW). F.D. is supported by GRAPPA PhD Fellowship. M.L. and I.S.Y. are supported in part by funding from the European Research Council under the European Union's Seventh Framework Programme (FP7/2007-2013) / ERC Grant agreement no.~268088-EMERGRAV.

\appendix


\section{Analytical details}
\label{sec-shift}

In this appendix, we will clarify some analytical details omitted in Sec.~\ref{sec-weakG}. There we showed how to reach a simple differential equation for $u_1$, Eq.~(\ref{eq-eomu1}), which can be solved simply by integrating the Green's function:
\begin{eqnarray}
u_1(r,t_f) &=& \frac{1}{2}\int_{-r_0}^{r_0}dt
\int_{r-r_0+t}^{r+r_0-t} dr'~S(r',t) 
 \nonumber \\
&=&  \frac{1}{2}\int_{-r_0}^{r_0}dt
\int_{r-r_0+t}^{r+r_0-t} dr'~ 
\left( C(\ddot{u}_0+u_0'') + \dot{C}\dot{u}_0+
C'\left(u_0'-\frac{u_0}{r'}\right) \right)~.
\label{eq-intu1}
\end{eqnarray}
Here we should be careful about our method of images. A physical solution $\phi_1$ is only given by a $u_1$ that is an odd function of $r$, and it is not obvious that the $u_1$ given by the above integral will have this property. Another potentially worrisome observation is that the lower limit of the $r'$ integral can be negative for some positive $r$, but a physical answer should only invoke an integration over the physical space $r>0$ where the quantity $C=V-M/r$ is naturally defined.

In this case, these concerns about the method of images can be easily resolved. As explained in Sec.~\ref{sec-weakG}, we can generalize the definition of $V$ and $M$ to include the $r<0$ region. We will find that $M$ is an odd function of $r$ and $V$ is even. Together with the fact that $u_0$ is odd, we see that the integrand in Eq.~(\ref{eq-intu1}) is odd. Any integration over negative $r$ is canceled by an equal region with positive $r$, so effectively the lower limit of the $r'$ integral is $|r-r_0+t|$. Eq.~(\ref{eq-intu1}) is effectively only integrating over the physical range. It is, however, more convenient to keep working in this form and avoid the confusion of taking an absolute value. An odd integrand here also guarantees that $u_1$ is an odd function which leads to a physical $\phi_1$.

The form of Eq.~(\ref{eq-intu1}) clearly suggests some integrations by parts.
\begin{eqnarray}
u_1(r,t_f) &=& -\frac{1}{2}\int_{-r_0}^{r_0}dt
\int_{r-r_0+t}^{r+r_0-t} dr'~
C'\frac{u_0}{r'}~ + ~\frac{1}{2}\int_{-r_0}^{r_0}dt
\int_{r-r_0+t}^{r+r_0-t} dr'~
(C u_0''+ C' u_0') 
\nonumber \\ \label{eq-long}
& &+
\frac{1}{2}\int_{r-2r_0}^{r+2r_0}dr'
\int_{-r_0}^{r_0-|r-r'|} dt~
(C\ddot{u}_0 + \dot{C}\dot{u}_0) \\ \nonumber
&=& -\frac{1}{2}\int_{-r_0}^{r_0}dt
\int_{r-r_0+t}^{r+r_0-t} dr'~C'\frac{u_0}{r'} \\ \nonumber
& &+ \frac{1}{2}\int_{-r_0}^{r_0}dt~
C\bigg[(r+r_0-t),t\bigg]u_0'\bigg[(r+r_0-t),t\bigg] \\ \nonumber
& &- \frac{1}{2}\int_{-r_0}^{r_0}dt~
C\bigg[(r-r_0+t),t\bigg]u_0'\bigg[(r-r_0+t),t\bigg] \\ \nonumber
& &+ \frac{1}{2}\int_{r-2r_0}^{r}dr'~
C\bigg[r',(r_0-r+r')\bigg] \dot{u}_0\bigg[r',(r_0-r+r')\bigg]
\\ \nonumber
& &+ \frac{1}{2}\int_{r}^{r+2r_0}dr'~
C\bigg[r',(r_0-r'+r)\bigg] \dot{u}_0\bigg[r',(r_0-r'+r)\bigg]
\\ \nonumber
& &- \frac{1}{2}\int_{r-2r_0}^{r+2r_0}dr'~
C(r',-r_0) \dot{u}_0(r',-r_0) 
\\ \nonumber
&=& ~\frac{1}{2}\int_{-r_0}^{r_0}dt~
C\bigg[(r-r_0+t),t\bigg]
\left(\dot{u}_0\bigg[(r-r_0+t),t\bigg] - u_0'\bigg[(r-r_0+t),t\bigg]\right) \\ \nonumber
& &+ \frac{1}{2}\int_{-r_0}^{r_0}dt~
C\bigg[(r+r_0-t),t\bigg]
\left(\dot{u}_0\bigg[(r+r_0-t),t\bigg] + u_0'\bigg[(r+r_0-t),t\bigg]\right) \\ \nonumber
& &-\frac{1}{2}\int_{-r_0}^{r_0}dt
\int_{r-r_0+t}^{r+r_0-t} dr'~C'\frac{u_0}{r'} 
-\frac{1}{2}\int_{r-2r_0}^{r+2r_0}dr'~
C(r',-r_0) \dot{u}_0(r',-r_0) ~.
\end{eqnarray}
In the above equation, we first isolated two terms which should be integrated by parts, and for one of them we interchange the order of integration so it can be done with respect to $t$ instead of $r'$. The integration by parts produces five boundary terms as line integral along five segments which we explicitly write down. Finally, two pairs of segments can combine with each other and be expressed as time integrals. We collect the remaining space integral and the only non-boundary term which cannot be integrated by parts in the end.

Note that up to this step, we have not used any approximations. We did not even use the property that $C$ is sourced by $\phi_0$. In other words, this expression could describe the change in the field profile under the influence of any other spherically symmetric gravitational effects, either apart from or on top of its self-interaction.

Our next step is to plug in Eq.~(\ref{eq-u0}) and use our assumption that it represents a thin shell: we assume that $u_0$ is only nonzero within two narrow packages around $r=t$ and $r=-t$. This significantly simplifies Eq.~(\ref{eq-long}) to
\begin{eqnarray}
\label{eq-u1}
u_1(r,t_f) &=& -\frac{\epsilon}{\sqrt{w}}~
f'\left(\frac{r-r_0}{w}\right) 
\int_{-r_0}^{r_0}dt~C\bigg[(r-r_0+t),t\bigg] \\ \nonumber
& & +\frac{\epsilon}{\sqrt{w}}~
f'\left(\frac{-r-r_0}{w}\right)
\int_{-r_0}^{r_0}dt~C\bigg[(r+r_0-t),t\bigg] \\ \nonumber
& &-\frac{1}{2}\int_{-r_0}^{r_0}dt
\int_{r-r_0+t}^{r+r_0-t} dr'~C'\frac{u_0}{r} 
-\frac{1}{2}\int_{r-2r_0}^{r+2r_0}dr'~
C(r',-r_0) \dot{u}_0(r',-r_0) ~.
\end{eqnarray}
Note that here the $f'$ means a derivative with respect to the variable of $f$ instead of a $r$ derivative, which should always be clear from the context.

Since in the end, we are only interested in the physical range $r>0$, we can actually drop the second term because the profile $f$ is zero there. This starts to take the promised form of Eq.~(\ref{eq-shellexp}), and we can almost identify
\begin{equation}
\Delta r = \int_{-r_0}^{r_0}dt~C\bigg[|r-r_0+t|,t\bigg]~.
\label{eq-shift}
\end{equation}
Note that we have added an absolute value to the first variable in $C$. This makes no difference since it is even, but it helps to emphasize the fact that the integral can be strictly limited to the physical $r>0$ region. 

The physical meaning of Eq.~(\ref{eq-shift}) now becomes clear. When the metric includes first-order corrections, such as in Eq.~(\ref{eq-metric1}), a null ray actually follows
\begin{equation}
\left(1+\frac{M}{r}\right)|dr| = (1+V)~dt~.
\end{equation} 
Thus an incoming null ray starting from $r=r_0$ and $t_i=-r_0$ does not exactly return to $r=r_0$ at $t_f=r_0$; the amount it misses is exactly given by Eq.~(\ref{eq-shift}).  The leading-order correction due to gravity, of course, includes the fact that geodesics are changed, and the shell simply follows the new geodesic. A geometric calculation is enough to determine how much a localized object appears to be shifted from the position predicted by the zeroth-order theory.

For any finite-sized source, the gravitational potential at large $r$ is proportional to $1/r$, so the integral in Eq.~(\ref{eq-shift}) actually had a piece proportional to $\log r_0$.  Since we are have taken $r_0$ to be large, one might have worried that such a term would the ruin perturbation expansion. However, such the $\log r_0$ term is totally expected from the change of geodesics and does not interfere with our goal of computing the change in width or other changes.

One last concern about the position shift in Eq.~(\ref{eq-shift}) is that it is a function of $r$. This turns out not to be a problem either, since the $r$-dependent part of $\Delta r$ is not proportional to $\log r_0$. We can see this by taking a derivative with respect to $r$:
\begin{eqnarray}
\partial_r\Delta r &=& \partial_r
\left( \int_{-r_0}^{r_0-r}dt~C\bigg[(r_0-r-t),t\bigg] + \int_{r_0-r}^{r_0}dt~C\bigg[(t-r_0+r),t\bigg] \right) \nonumber \\
&=& \left( -\int_{-r_0}^{r_0-r}dt~C'\bigg[(r_0-r-t),t\bigg] + \int_{r_0-r}^{r_0}dt~C'\bigg[(t-r_0+r),t\bigg] \right)~.
\label{eq-shiftvary}
\end{eqnarray}
According to the Einstein's equation, we have
\begin{equation}
C' = V'-\frac{M'}{r}+\frac{M}{r^2} 
= \frac{2M}{r^2} + \frac{r}{2}\left(T_{rr}-T_{tt}\right)~.
\label{eq-C'}
\end{equation}
This means that as long as we restrict the matter sources to (1) finite-sized sources that vanish beyond some fixed $r$ and/or (2) radiation in the radial direction, $T_{rr}=T_{tt}$, then the $r$ dependence of $\Delta r$ will not have a $\log  r_0$ (or any other large $r_0$) dependence.  Furthermore, there is no small-$r$ divergence either, since the two terms in Eq.~(\ref{eq-shiftvary}) takes opposite signs and cancel each other near $r=0$. Pictorially, this means that different infinitesimal segments within the wavepacket ``shift'' differently from one another by some finite amount.

In the last line of Eq.~(\ref{eq-u1}), the first term is also finite for the same reason as Eq.~(\ref{eq-C'}), and the second term is finite because $u_0$ has compact support. These terms should be combined and understood as some perturbative deformations of the wavepacket profile. They are cleanly separated from the $\Delta r\sim\log r_0$ overall shift, which is uniquely defined by a projection:
\begin{equation}
\Delta r = -\frac{\int u_1\partial_ru_0~dr}{\int (\partial_ru_0)^2~dr}~.
\label{eq-dr}
\end{equation}
We can simply remove this shift mode from Eq.~(\ref{eq-shellexp}) and study the other deformations. A more physical way to understand the removal of this shift is letting the wavepacket evolve an extra time $\Delta t=\Delta r$ such that it really reaches position $r_0$; then it will be fair to compare with the zeroth-order profile at the same position.

In order to eventually form a black hole, we need the energy density to become large. Since the total energy is conserved, the most trivial way to increase the energy density is to narrow the width of the profile. The leading-order change in width can be extracted from $u_1$ by the following projection:
\begin{equation}
\Delta w = 
\frac{\int u_1\partial_wu_0~dr}{\int (\partial_wu_0)^2~dr}~.
\label{eq-dw}
\end{equation}
Note that the $\epsilon$ dependence was already scaled out in Eq.~(\ref{eq-scale}). Interestingly, $\epsilon^2$ has the unit of length in our conventions, and the physical change in width is $\epsilon^2\Delta w$. Therefore, $\Delta w$ is dimensionless. The width $w$ is the only other dimensionful quantity that can potentially affect $\Delta w$ in the leading order ($r_0$ affects only the subleading error), and so there is no way it can enter the expression for $\Delta w$.

What we really wish to determine the sign of $\Delta w$. $\Delta w<0$ means that the shell gets narrower, and several bounces later it might form a black hole. On the other hand, $\Delta w>0$ means that the shell gets wider, and several bounces later the energy will be more diluted, which in some sense is moving away from a black hole in phase space.

One technical point to note here is that, given our shell profile Eq.~(\ref{eq-u0}), the mode $\partial_wu_0$ actually measures the scaling the profile around some center. If that center is not the center of mass, then this scaling not only changes the width but also shifts the position. A simple projection will be contaminated by the large $\sim\log r_0$ contribution from the position shift. We will avoid this by always defining the profile $f(x)$ to have its center of mass at $x=0$. This means that, on top of the normalization
\begin{equation}
\int f'(x)^2~dx = 1~,
\end{equation}
we also demand that
\begin{equation}
\int xf'(x)^2~dx = 0~.
\end{equation}
This guarantees that the scaling mode $\partial_wu_0$ is orthogonal to the shift mode $\partial_ru_0$.


\section{Numerical examples}
\label{sec-examples}

\subsection{The asymmetry-focusing correlation}

In this appendix, we numerically evaluate the change to a thin-shell profile after one bounce. Our example will be the following two one-parameter families of profiles.
\begin{eqnarray}
g_a(x) &=& (1+ax \pm x^2)~e^{-x^2}~, \label{eq-profilea1} \\
N_a &=& \sqrt{\int g_a'^2~dx}~, \label{eq-profilea2} \\
X_a &=& N_a^{-2}\int xg_a'^2~dx~,\label{eq-profilea3} \\
f_a(x) &\equiv& N_a^{-1}g_a(x+X_a)~. \label{eq-profilea4}
\end{eqnarray}
They are symmetric when $a=0$, and varying $a$ scans through two different directions of asymmetry. Note that the quadratic term is necessary. Without it, a small $a$ simply means an overall shift in position and the profile will be still symmetric to leading order. Our definition of $f_a(x)$ shifts the center of mass back to $x=0$ and preserves only the asymmetry generated by $a$.

\begin{figure}[tb]
\centering
\begin{subfigure}{.45\textwidth}
  \centering
  \includegraphics[width=.8\linewidth]{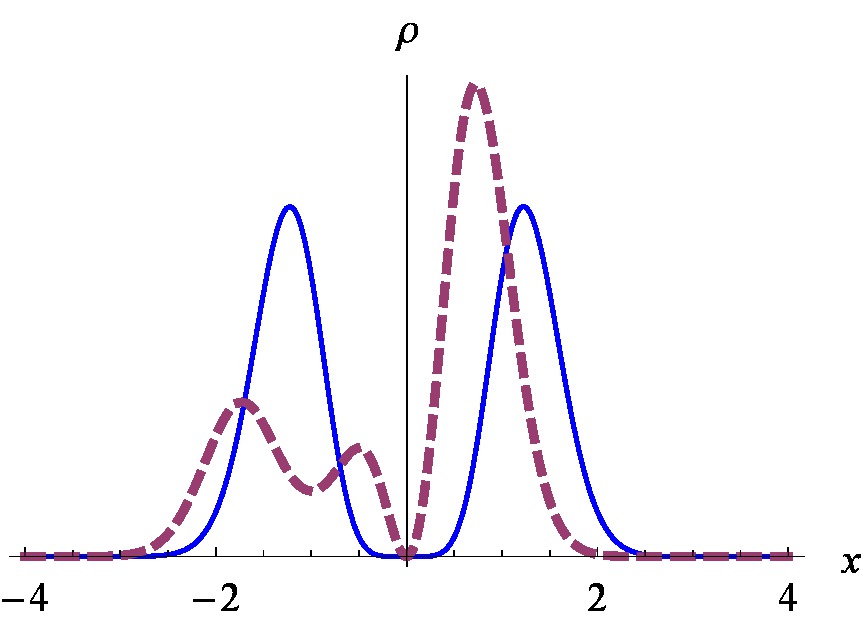}
\end{subfigure}%
\begin{subfigure}{.45\textwidth}
  \centering
  \includegraphics[width=.8\linewidth]{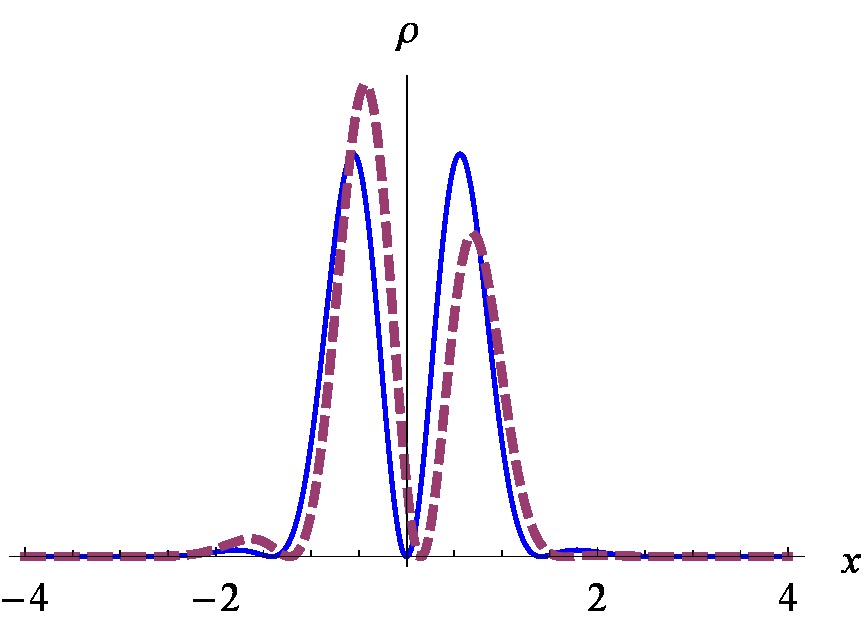}
\end{subfigure}
\caption{The left panel shows the energy density of the profiles with the ``$+$'' sign defined in Eq.~(\ref{eq-profilea1}), and the right panel shows the profiles with the ``$-$'' sign. The blue curves are the symmetric, $a=0$ profiles. The red (dashed) curves are for $a=0.5$, which is the maximally ``tilted'' profile in the range we scanned through. We can see that the family with the ``+'' sign is more sensitive to the parameter $a$.}
\label{fig-profiles}
\end{figure}

We plot some representative profiles in Fig.~\ref{fig-profiles}. Note that for both families, we have $f_a(x)=f_{-a}(-x)$. Scanning through positive and negative values of $a$ can confirm our analytical proof in Sec.~\ref{sec-proof} that flipping the profiles leads to opposite behaviors within one bounce. It will also provide a better understanding about what physical quantity really affects whether a profile becomes focused or not.

There are infinite ways to be asymmetric, and our parameter $a$ certainly is not the unique parameter to quantify the asymmetry. It also has no reason to be the asymmetry directly responsible for focusing or defocusing the profile. However, for any family of profiles centered around a symmetric one, we can define a natural parameter to quantify the asymmetry, at least for small values of $a$. Here is how it goes. First of all, $g_a$ has a center of mass shifted by $X_a$ from $g_0$ by the definition in Eq.~(\ref{eq-profilea3}). On the other hand, one can also naturally define the shift by a projection to the zero mode, which is exactly the way we defined $\Delta r$ in Eq.~(\ref{eq-dr}).
\begin{equation}
\bar{X}_a = N_0^{-2} \int [g_0(x)-g_a(x)]g'_0(x)~dx~.
\end{equation}
These two shifts already disagree at linear order in $a$, therefore the amount of their disagreement, $\Delta X_a = (X_a-\bar{X}_a)$, seems to be a reasonable way to quantify the asymmetry.

\begin{figure}[tb]
\centering
\begin{subfigure}{.45\textwidth}
  \centering
  \includegraphics[width=.8\linewidth]{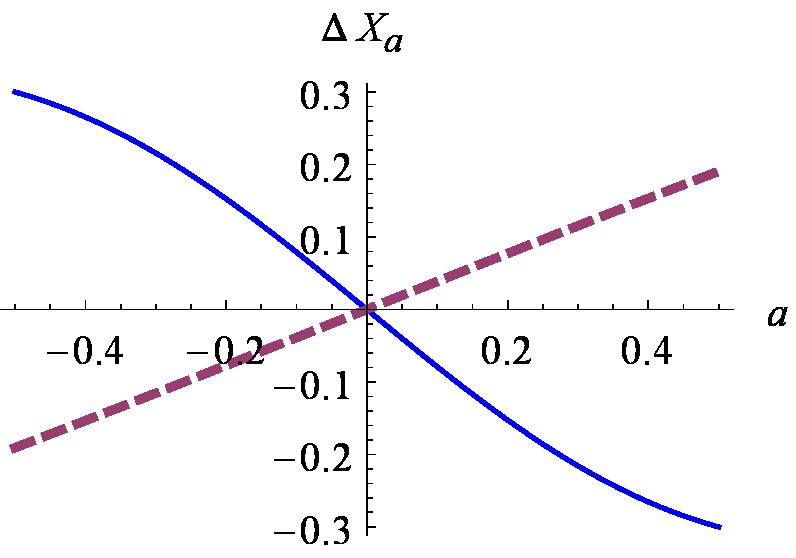}
\end{subfigure}%
\begin{subfigure}{.45\textwidth}
  \centering
  \includegraphics[width=.8\linewidth]{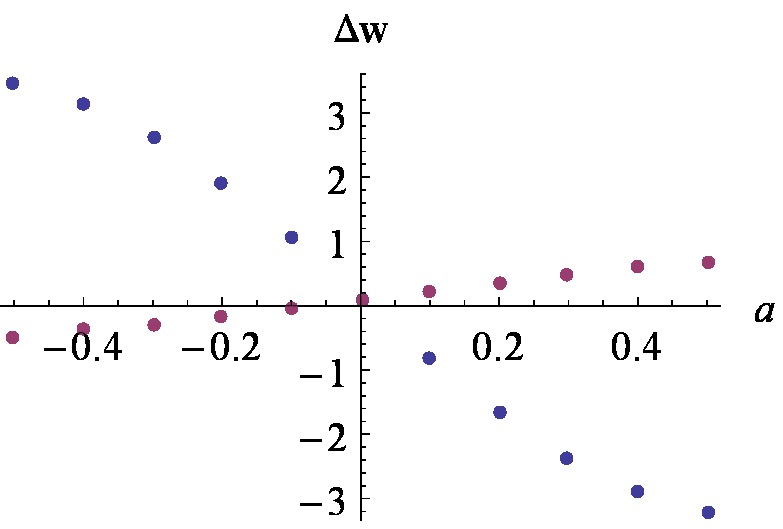}
\end{subfigure}
\caption{The left panel shows the asymmetry parameter defined as the difference between the center-of-mass shift and the field-profile shift. The blue curve is for the ``$+$'' family and the red curve for the ``$-$'' family. The right panel shows the change is width after one bounce, which qualitatively agrees with the asymmetry parameter. These are done with $r_0=60$ and $w=1$. Recall that the physical change in width is actually $\epsilon^2 \Delta w$.}
\label{fig-ScanShape}
\end{figure}

Given these profiles, we solve Eq.~(\ref{eq-metric1}) for $M$ and $V$, and then we can integrate Eq.~(\ref{eq-intu1}). When we scan the parameter $a$ from $-0.5$ to $0.5$, the change in width $\Delta w$ defined in Eq.~(\ref{eq-dw}) follows a pattern closely resembling the behavior of this asymmetry parameter, $\Delta X_a$. We compare them side-by-side in Fig.~\ref{fig-ScanShape}. Note that they are not identical. For example, the relative slopes between the two families near $a=0$ are not the same. Thus, although we see a rough correlation between them, we cannot claim that our asymmetry parameter directly controls focusing or widening in one bounce. 

In our conventions, the profiles are moving toward the right in Fig.~\ref{fig-profiles}. If we compare their shapes in Fig.~\ref{fig-profiles} to their behaviors in Fig.~\ref{fig-ScanShape}, we get the following impression:
\begin{itemize}
\item When the wavepacket is denser in the front, we get $\Delta w<0$. The shell gets focused into a smaller region, and gravitational effect will become stronger in the next bounce. 
\item When the wavepacket is denser in the tail, we get $\Delta w>0$. It profile gets wider after one bounce, and gravitational effect will become weaker in the next bounce.
\end{itemize}
Since the family of profiles with ``$+$'' sign is much more sensitive to the parameter $a$, we will use it to test other behaviors later in App.~\ref{sec-AnotherObject}.

In Fig.~\ref{fig-ScanShape}, one might notice that for the $a=0$ symmetric profiles, the changes in width are not exactly zero as we argued earlier. This deviation is not physical but simply an artifact of our approximation. That is because although the physical solution is symmetric in time, our technical choice breaks that symmetry by a small amount. We have set the correction to the field profile at the initial time to zero, $u_1(r, -r_0)=0$. This is a small error since the first-order correction to the metric in Eq.~(\ref{eq-metric1}) would have already modified the free field profile at that time, by a small fraction $\sim \epsilon^2V \sim \epsilon^2/r_0$.

\begin{figure}[tb]
\centering
\begin{subfigure}{.45\textwidth}
  \centering
  \includegraphics[width=.8\linewidth]{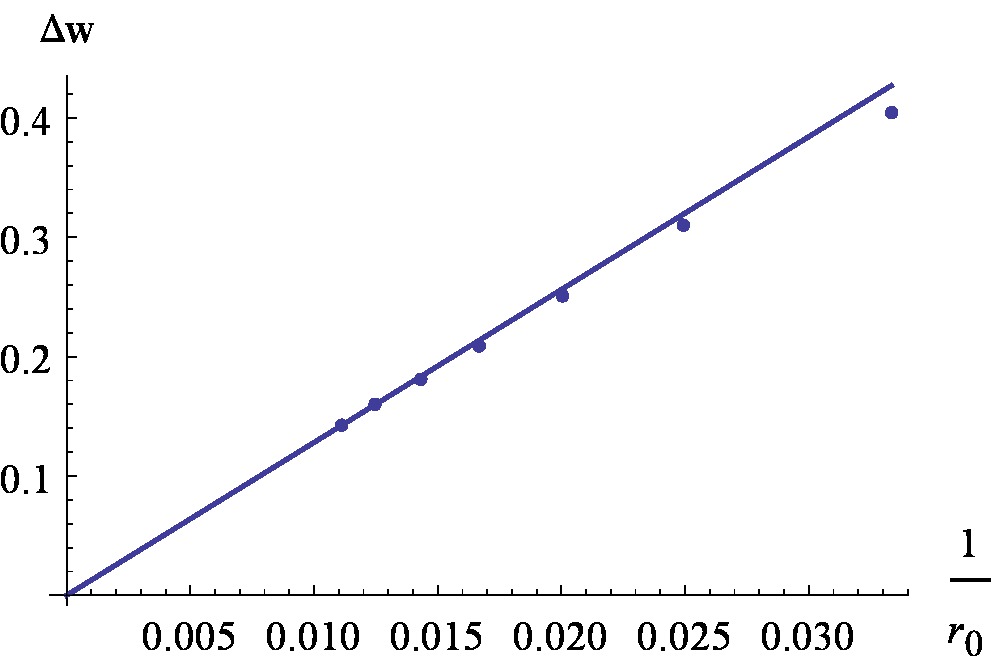}
\end{subfigure}%
\begin{subfigure}{.45\textwidth}
  \centering
  \includegraphics[width=.8\linewidth]{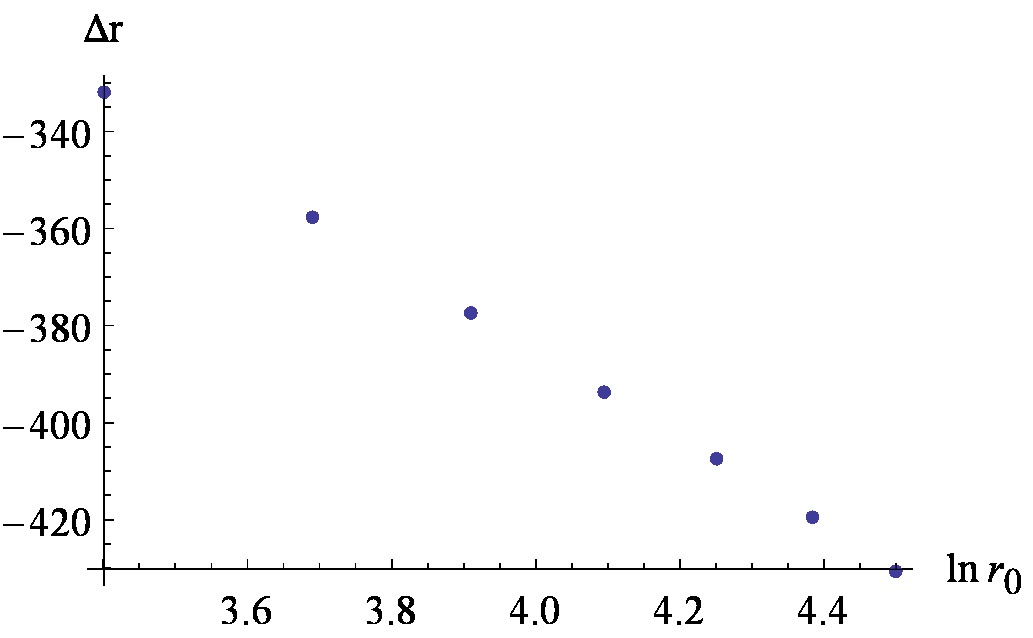}
\end{subfigure}
\caption{Results with $a=0$ and varying $r_0$ from 30 to 90 in steps of 10. The left panel shows $\Delta w$, which indeed goes to zero as $1/r_0$. The right panel shows the position shift $\Delta r$ defined in Eq.~(\ref{eq-dr}) which has the correct $\log r_0$ dependence.
}
\label{fig-r0}
\end{figure}

We test this explanation by varying $r_0$ and verifying that $\Delta w$ goes to zero in the expected way; see Fig.~\ref{fig-r0}. We also verified that the position shift indeed has an $r_0$-dependent shift $\Delta r\propto \log r_0$, as discussed in App.~\ref{sec-shift}.

Finally, with a symmetric profile, one can ask for a prediction for the next bounce. What we observed in these examples is that a symmetric profile will pick up a $\Delta X_a>0$ in one bounce. This is a very tentative evidence that in the next bounce, they will have $\Delta w>0$, namely their energy become defocused. We stress again that this is not a proof, but merely two examples we observed. A more thorough investigation is required to support the generic downward flow we conjectured in Sec.\ref{sec-dis}.

\subsection{The effect of another object}
\label{sec-AnotherObject}

Black hole formation does not always involve all of the energy becoming concentrated into a thin shell. For example, an initially smooth field profile might start to develop one or more sharp peaks. It is possible for the energy density in these peaks to become large enough to induce strong gravity and black hole collapse before or even without the average density of the entire profile ever becoming large.

In this section, we will present some evidence that sharp peaks evolve similarly to thin shells.  In the perturbative regime, one is free to separate the matter into components in many ways. In particular, we can treat a smooth field profile with a sharp peak as a thin shell propagating in the background of some additional diffuse source of gravity. Our approach is convenient since Eq.~(\ref{eq-intu1}) and further analysis about it do not rely on the specific metric ansatz Eq.~(\ref{eq-metric1}). As long as the additional source are also spherically symmetric, we can simply repeat the calculation in the previous appendix.

We will start by considering a simple situation in which the additional matter sources are static. In addition to the thin shell with total mass $4\pi\epsilon^2$, we have
\begin{eqnarray}
M_0(r) &=& 10\epsilon_{\rm star}^2
\tanh \left(\frac{r}{w_{\rm star}}\right)^2~, \\
P_r &\approx& 0~.
\end{eqnarray}
This is a star of roughly constant density, radius $w_{\rm star}$ and total mass $10\epsilon_{\rm star}^2$. It does not interact with the massless field which forms the shell in any other way other than gravitationally, so it simply enters  by altering the metric in Eq.~(\ref{eq-metric1}). We assume the star is stable and supports itself by a negligible amount of radial pressure (but it can have angular pressure), so it does not add any extra term to modify $T_{rr}$.

According to the momentum space analysis, including this additional gravitational source breaks the AdS resonance structure and should interfere with black hole formation \cite{MalRos14}. We show that such an interpretation is not necessary to understand the dynamics of thin shells in one bounce. Remember that for a symmetric shell profile, we argued that there cannot be a change in width due to the time-reversal symmetry.  
 Adding an extra, static source does not break that symmetry, so symmetric profiles again cannot change in size. And, it is straightforward to verify that asymmetry still focuses or defocuses in qualitatively the same way as before. 

\begin{figure}[tb]
\centering
\begin{subfigure}{.45\textwidth}
  \centering
  \includegraphics[width=.8\linewidth]{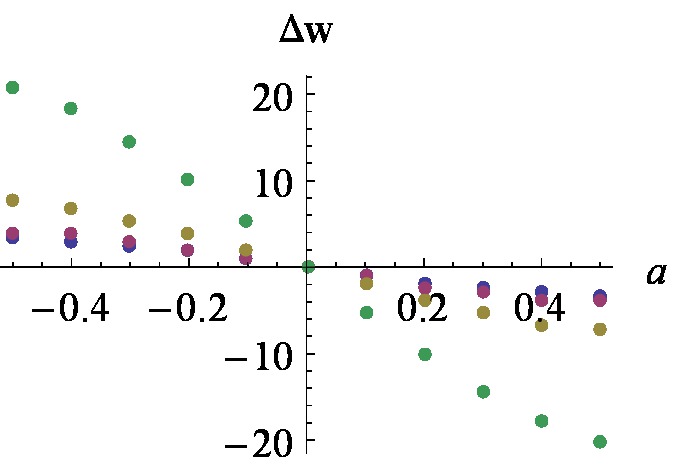}
\end{subfigure}%
\begin{subfigure}{.45\textwidth}
  \centering
  \includegraphics[width=.8\linewidth]{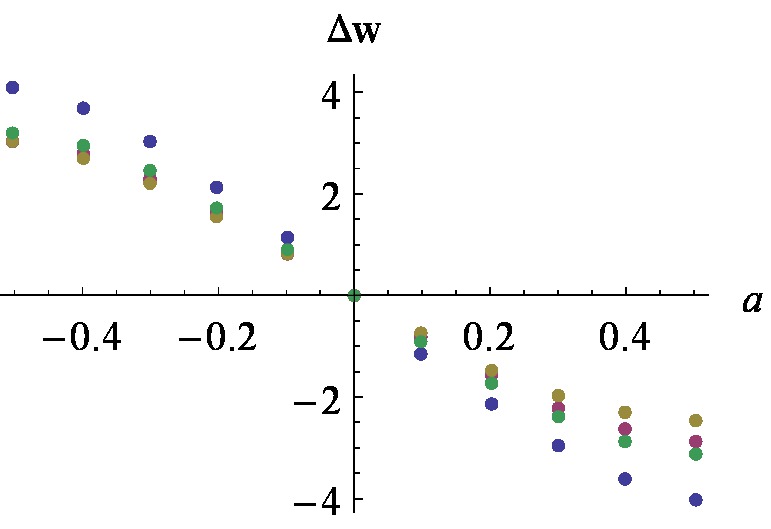}
\end{subfigure}
\caption{We plot the change in width $\Delta w$ as a function of the parameter $a$ in Eq.~(\ref{eq-profilea1}). Both figures are with $r_0=30$. The left panel shows the effect of dialing the mass of the additional matter source, $\epsilon_{\rm star}^2=0,10,50,200$, without changing its size $w_{\rm star}=1$. The right panel shows the effect of dialing the size while keeping the same density, $w_{\rm star}=1,5,20,200$. We have removed the errant $1/r_0$ contribution by hand.}
\label{fig-addMass}
\end{figure}

In Fig.~\ref{fig-addMass}, we demonstrate that whether the shell gets thinner or thicker has the same dependence on the asymmetry induced by the parameter $a$. Its magnitude does have an interesting dependence on the additional source. In the first set of data we fix the size of the star to be equal to the shell. The change in width turns out to grow linearly with the additional mass. On the other hand if we keep the same density and increase the size of the star, the change in width is not strongly affected. 

\bibliographystyle{utcaps}
\bibliography{all}

\end{document}